\input epsf
\input amssym
\input eplain
\catcode`\@11\relax
\newif\ify@autoscale \y@autoscaletrue \def\Yautoscale#1{\ifnum #1=0
  \y@autoscalefalse\else\y@autoscaletrue\fi}
\newdimen\y@b@xdim
\newdimen\y@boxdim \y@boxdim=13pt
\def\Yboxdim#1{\y@autoscalefalse\y@boxdim=#1}
\newdimen\y@linethick    \y@linethick=.3pt
\def\Ylinethick#1{\y@linethick=#1}
\newskip\y@interspace \y@interspace=0ex plus 0.3ex
\def\Yinterspace#1{\y@interspace=#1}
\newif\ify@vcenter   \y@vcenterfalse
\def\Yvcentermath#1{\ifnum #1=0 \y@vcenterfalse\else\y@vcentertrue\fi}
\newif\ify@stdtext   \y@stdtextfalse
\def\Ystdtext#1{\ifnum #1=0 \y@stdtextfalse\else\y@stdtexttrue\fi}
\newif\ify@enable@skew   \y@enable@skewfalse
\expandafter\ifx\csname enableskew\endcsname\relax
 \y@enable@skewfalse \else \y@enable@skewtrue\fi
\def\y@vr{\vrule height0.8\y@b@xdim width\y@linethick depth 0.2\y@b@xdim}
\def\y@emptybox{\y@vr\hbox to \y@b@xdim{\hfil}}
\ify@enable@skew
 \def\y@abcbox#1{\if :#1\else
   \y@vr\hbox to \y@b@xdim{\hfil#1\hfil}\fi}
 \def\y@mathabcbox#1{\if :#1\else
   \y@vr\hbox to \y@b@xdim{\hfil$#1$\hfil}\fi}
\else
 \def\y@abcbox#1{\y@vr\hbox to \y@b@xdim{\hfil#1\hfil}}
 \def\y@mathabcbox#1{\y@vr\hbox to \y@b@xdim{\hfil$#1$\hfil}}
\fi
\def\y@setdim{%
  \ify@autoscale%
   \ifvoid1\else\typeout{Package youngtab: box1 not free! Expect an
     error!}\fi%
   \setbox1=\hbox{A}\y@b@xdim=1.6\ht1 \setbox1=\hbox{}\box1%
  \else\y@b@xdim=\y@boxdim \advance\y@b@xdim by -2\y@linethick
  \fi}
\newcount\y@counter
\newif\ify@islastarg
\def\y@lastargtest#1,#2 {\if\space #2 \y@islastargtrue
  \else\y@islastargfalse\fi}
\def\y@emptyboxes#1{\y@counter=#1\loop\ifnum\y@counter>0
  \advance\y@counter by -1 \y@emptybox\repeat}
\def\y@nelineemptyboxes#1{%
  \vbox{%
    \hrule height\y@linethick%
    \hbox{\y@emptyboxes{#1}\y@vr}
    \hrule height\y@linethick}\vskip-\y@linethick}
\def\yng(#1){%
  \y@setdim%
  \hskip\y@interspace%
  \ifmmode\ify@vcenter\vcenter\fi\fi{%
  \y@lastargtest#1,
  \vbox{\offinterlineskip
    \ify@islastarg
     \y@nelineemptyboxes{#1}
    \else
     \y@ungempty(#1)
    \fi}}\hskip\y@interspace}
\def\y@ungempty(#1,#2){%
  \y@nelineemptyboxes{#1}
  \y@lastargtest#2,
  \ify@islastarg
   \y@nelineemptyboxes{#2}
  \else
   \y@ungempty(#2)
  \fi}
\def\y@nelettertest#1#2. {\if\space #2 \y@islastargtrue
  \else\y@islastargfalse\fi}
\def\y@abcboxes#1#2.{%
  \ify@stdtext\y@abcbox#1\else\y@mathabcbox#1\fi%
  \y@nelettertest #2.
  \ify@islastarg\unskip%
   \ify@stdtext\y@abcbox{#2}\else\y@mathabcbox{#2}\fi%
  \else\y@abcboxes#2.\fi}
 \newdimen\y@full@b@xdim
 \newcount\y@m@veright@cnt
\ify@enable@skew
 \def\y@get@m@veright@cnt#1#2.{%
   \if :#1 \advance\y@m@veright@cnt by 1\y@get@m@veright@cnt#2.\fi}
 \let\y@setdim@=\y@setdim
 \def\y@setdim{%
   \y@setdim@ \y@full@b@xdim=\y@b@xdim
   \advance\y@full@b@xdim by 1\y@linethick}
 \def\y@m@veright@ifskew#1{
   \y@m@veright@cnt=0 \y@get@m@veright@cnt#1.
   \moveright \y@m@veright@cnt\y@full@b@xdim}
\else
 \def\y@m@veright@ifskew#1{}
\fi
\def\y@nelineabcboxes#1{%
  \y@nelettertest #1.
  \ify@islastarg
   \y@m@veright@ifskew{#1}
    \vbox{
      \hrule height\y@linethick%
      \hbox{\ify@stdtext\y@abcbox#1\else\y@mathabcbox#1\fi\y@vr}
      \hrule height\y@linethick}\vskip-\y@linethick
  \else
   \y@m@veright@ifskew{#1}
    \vbox{
      \hrule height\y@linethick%
      \hbox{\y@abcboxes #1.\y@vr}%
      \hrule height\y@linethick}\vskip-\y@linethick
  \fi}
\def\young(#1){%
  \y@setdim%
  \hskip\y@interspace%
  \y@lastargtest#1,
  \ifmmode\ify@vcenter\vcenter\fi\fi{%
  \vbox{\offinterlineskip
    \ify@islastarg\y@nelineabcboxes{#1}%
    \else\y@ungabc(#1)%
    \fi}}\hskip\y@interspace}
\def\y@ungabc(#1,#2){%
  \y@nelineabcboxes{#1}%
  \y@lastargtest#2,
  \ify@islastarg\y@nelineabcboxes{#2}%
  \else\y@ungabc(#2)%
  \fi}
\catcode`\@12\relax
 

\input\jobname.intref

\newfam\scrfam
\batchmode\font\tenscr=rsfs10 \errorstopmode
\ifx\tenscr\nullfont
        \message{rsfs script font not available. Replacing with calligraphic.}
        \def\scr{\cal}
\else   
        \font\sevenscr=rsfs7
        \font\fivescr=rsfs5
        \skewchar\tenscr='177 \skewchar\sevenscr='177 \skewchar\fivescr='177
        \textfont\scrfam=\tenscr \scriptfont\scrfam=\sevenscr
        \scriptscriptfont\scrfam=\fivescr
        \def\scr{\fam\scrfam}
        \def\cal{\scr}
\fi
\catcode`\@=11
\newfam\frakfam
\batchmode\font\tenfrak=eufm10 \errorstopmode
\ifx\tenfrak\nullfont
        \message{eufm font not available. Replacing with italic.}
        
\else
	
	\font\sevenfrak=eufm7 \font\fivefrak=eufm5
        
	\textfont\frakfam=\tenfrak
	\scriptfont\frakfam=\sevenfrak \scriptscriptfont\frakfam=\fivefrak
	
\fi
\catcode`\@=\active
\newfam\msbfam
\batchmode\font\twelvemsb=msbm10 scaled\magstep1 \errorstopmode
\ifx\twelvemsb\nullfont\def\Bbb{\bf}
        
	\font\eightbbb=cmb10 at 8pt
	\message{Blackboard bold not available. Replacing with boldface.}
\else   \catcode`\@=11
        \font\tenmsb=msbm10 \font\sevenmsb=msbm7 \font\fivemsb=msbm5
        \textfont\msbfam=\tenmsb
        \scriptfont\msbfam=\sevenmsb \scriptscriptfont\msbfam=\fivemsb
        \def\Bbb{\relax\expandafter\Bbb@}
        \def\Bbb@#1{{\Bbb@@{#1}}}
        \def\Bbb@@#1{\fam\msbfam\relax#1}
        \catcode`\@=\active
	
	\font\eightbbb=msbm8
\fi
        \font\fivemi=cmmi5
        \font\sixmi=cmmi6
        \font\eightrm=cmr8              \def\xrm{\eightrm}
        \font\eightbf=cmbx8             \def\xbf{\eightbf}
        \font\eightit=cmti10 at 8pt     \def\xit{\eightit}
        \font\sixit=cmti8 at 6pt        
        
        \font\eighttt=cmtt8             
        \font\sixtt=cmtt6
        
        \font\eightcp=cmcsc8    
                      \def\xold{\eighti}
        \font\eightmi=cmmi8
                     \def\xbold{\eightib}
        \font\teni=cmmi10               \def\old{\teni}
        \font\tencp=cmcsc10

        \font\twelvecp=cmcsc10 scaled\magstep1
        
        \font\sixrm=cmr6
        \font\fiverm=cmr5

        \font\eightsy=cmsy8
        \font\sixsy=cmsy6
        \font\eightsl=cmsl8
        \font\sixsl=cmsl6
        
        \font\sixbf=cmbx6

	 at10pt	
	\font\twelvehelvbold=phvb at12pt
	 at14pt
	\font\sixteenhelvbold=phvb at16pt
	 at16pt



\def\xbold{\xbf}
\def\xold{\xrm}


\def\noblackbox{\overfullrule=0pt}
\noblackbox

\def\eightpoint{
\def\rm{\fam0\eightrm}
\textfont0=\eightrm \scriptfont0=\sixrm \scriptscriptfont0=\fiverm
\textfont1=\eightmi  \scriptfont1=\sixmi  \scriptscriptfont1=\fivemi
\textfont2=\eightsy \scriptfont2=\sixsy \scriptscriptfont2=\fivesy
\textfont3=\tenex   \scriptfont3=\tenex \scriptscriptfont3=\tenex
\textfont\itfam=\eightit \def\it{\fam\itfam\eightit}
\textfont\slfam=\eightsl \def\sl{\fam\slfam\eightsl}
\textfont\ttfam=\eighttt \def\tt{\fam\ttfam\eighttt}
\textfont\bffam=\eightbf \scriptfont\bffam=\sixbf 
                         \scriptscriptfont\bffam=\fivebf
                         \def\bf{\fam\bffam\eightbf}
\normalbaselineskip=10pt}

\def\sixpoint{
\def\rm{\fam0\sixrm}
\textfont0=\sixrm \scriptfont0=\fiverm \scriptscriptfont0=\fiverm
\textfont1=\sixmi  \scriptfont1=\fivemi  \scriptscriptfont1=\fivemi
\textfont2=\sixsy \scriptfont2=\fivesy \scriptscriptfont2=\fivesy
\textfont3=\tenex   \scriptfont3=\tenex \scriptscriptfont3=\tenex
\textfont\itfam=\sixit \def\it{\fam\itfam\sixit}
\textfont\slfam=\sixsl \def\sl{\fam\slfam\sixsl}
\textfont\ttfam=\sixtt \def\tt{\fam\ttfam\sixtt}
\textfont\bffam=\sixbf \scriptfont\bffam=\fivebf 
                         \scriptscriptfont\bffam=\fivebf
                         \def\bf{\fam\bffam\sixbf}
\normalbaselineskip=8pt}


\newtoks\headtext
\headline={\ifnum\pageno=1\hfill\else
	\ifodd\pageno{\eightcp\the\headtext}{ }\dotfill{ }{\old\folio}
	\else{\old\folio}{ }\dotfill{ }{\eightcp\the\headtext}\fi
	\fi}
\def\makeheadline{\vbox to 0pt{\vss\noindent\the\headline\break
\hbox to\hsize{\hfill}}
        \vskip2\baselineskip}
\newcount\infootnote
\infootnote=0
\newcount\footnotecount
\footnotecount=1
\def\foot#1{\infootnote=1
\footnote{${}^{\the\footnotecount}$}{\vtop{\baselineskip=.75\baselineskip
\advance\hsize by
-\parindent{\eightpoint\rm\hskip-\parindent
#1}\hfill\vskip\parskip}}\infootnote=0\global\advance\footnotecount by
1}
\newcount\refcount
\refcount=1
\newwrite\refwrite
\def\oldsize{\ifnum\infootnote=1\xold\else\old\fi}
\def\ref#1#2{
	\def#1{{{\oldsize\the\refcount}}\ifnum\the\refcount=1\immediate\openout\refwrite=\jobname.refs\fi\immediate\write\refwrite{\item{[{\xold\the\refcount}]} 
	#2\hfill\par\vskip-2pt}\xdef#1{{\noexpand\oldsize\the\refcount}}\global\advance\refcount by 1}
	}
\def\refout{\eightpoint\catcode`\@=11
        \xrm\immediate\closeout\refwrite
        \vskip2\baselineskip
        {\noindent\twelvecp References}\hfill\vskip\baselineskip
        \baselineskip=.75\baselineskip
        \input\jobname.refs
        \baselineskip=4\baselineskip \divide\baselineskip by 3
        \catcode`\@=\active\rm}

\def\skipref#1{\hbox to15pt{\phantom{#1}\hfill}\hskip-15pt}

\def\hepth#1{\href{http://xxx.lanl.gov/abs/hep-th/#1}{arXiv:\allowbreak
hep-th\slash{\xold#1}}}

\def\arxiv#1#2{\href{http://arxiv.org/abs/#1.#2}{arXiv:\allowbreak
{\xold#1}.{\xold#2}}} 
 
\def\jhep#1#2#3#4{\href{http://jhep.sissa.it/stdsearch?paper=#2\%28#3\%29#4}{J. High Energy Phys. {\xbold #1#2} ({\xold#3}) {\xold#4}}}

\def\FP#1#2#3{Fortsch. Phys. {\xbold#1} ({\xold#2}) {\xold#3}}

\def\JHEP{\jhep}

\def\JPA#1#2#3{J. Phys. {\xbf A}{\xbold#1} ({\xold#2}) {\xold#3}}

\def\MPLA#1#2#3{Mod. Phys. Lett. {\xbf A}{\xbold#1} ({\xold#2}) {\xold#3}}

\def\NPB#1#2#3{Nucl. Phys. {\xbf B}{\xbold#1} ({\xold#2}) {\xold#3}}

\def\PLB#1#2#3{Phys. Lett. {\xbf B}{\xbold#1} ({\xold#2}) {\xold#3}}

\def\PRD#1#2#3{Phys. Rev. {\xbf D}{\xbold#1} ({\xold#2}) {\xold#3}}
\def\PRL#1#2#3{Phys. Rev. Lett. {\xbold#1} ({\xold#2}) {\xold#3}}

\newcount\sectioncount
\sectioncount=0
\def\section#1#2{\global\eqcount=0
	\global\subsectioncount=0
        \global\advance\sectioncount by 1
	\ifnum\sectioncount>1
	        \vskip2\baselineskip
	\fi
\noindent{\twelvecp\the\sectioncount. #2}\par\nobreak
       \vskip.5\baselineskip\noindent
        \xdef#1{{\old\the\sectioncount}}}
\newcount\subsectioncount
\def\subsection#1#2{\global\advance\subsectioncount by 1
\vskip.75\baselineskip\noindent\line{\tencp\the\sectioncount.\the\subsectioncount. #2\hfill}\nobreak\vskip.4\baselineskip\nobreak\noindent\xdef#1{{\old\the\sectioncount}.{\old\the\subsectioncount}}}
\def\immediatesubsection#1#2{\global\advance\subsectioncount by 1
\vskip-\baselineskip\noindent
\line{\tencp\the\sectioncount.\the\subsectioncount. #2\hfill}
	\vskip.5\baselineskip\noindent
	\xdef#1{{\old\the\sectioncount}.{\old\the\subsectioncount}}}
\newcount\subsubsectioncount
\def\subsubsection#1#2{\global\advance\subsubsectioncount by 1
\vskip.75\baselineskip\noindent\line{\tencp\the\sectioncount.\the\subsectioncount.\the\subsubsectioncount. #2\hfill}\nobreak\vskip.4\baselineskip\nobreak\noindent\xdef#1{{\old\the\sectioncount}.{\old\the\subsectioncount}.{\old\the\subsubsectioncount}}}
\newcount\appendixcount
\appendixcount=0
\def\appendix#1{\global\eqcount=0
        \global\advance\appendixcount by 1
        \vskip2\baselineskip\noindent
        \ifnum\the\appendixcount=1
        {\twelvecp Appendix A: #1}\par\nobreak
                        \vskip.5\baselineskip\noindent\fi
        \ifnum\the\appendixcount=2
        {\twelvecp Appendix B: #1}\par\nobreak
                        \vskip.5\baselineskip\noindent\fi
        \ifnum\the\appendixcount=3
        {\twelvecp Appendix C: #1}\par\nobreak
                        \vskip.5\baselineskip\noindent\fi}
\def\acknowledgements{\vskip2\baselineskip\noindent
        \underbar{\it Acknowledgements:}\ }
\newcount\eqcount
\eqcount=0
\def\Eqn#1{\global\advance\eqcount by 1
\ifnum\the\sectioncount=0
	\xdef#1{{\noexpand\oldsize\the\eqcount}}
	\eqno({\oldstyle\the\eqcount})
\else
        \ifnum\the\appendixcount=0
\xdef#1{{\noexpand\oldsize\the\sectioncount}.{\noexpand\oldsize\the\eqcount}}
                \eqno({\oldstyle\the\sectioncount}.{\oldstyle\the\eqcount})\fi
        \ifnum\the\appendixcount=1
	        \xdef#1{{\noexpand\oldstyle A}.{\noexpand\oldstyle\the\eqcount}}
                \eqno({\oldstyle A}.{\oldstyle\the\eqcount})\fi
        \ifnum\the\appendixcount=2
	        \xdef#1{{\noexpand\oldstyle B}.{\noexpand\oldstyle\the\eqcount}}
                \eqno({\oldstyle B}.{\oldstyle\the\eqcount})\fi
        \ifnum\the\appendixcount=3
	        \xdef#1{{\noexpand\oldstyle C}.{\noexpand\oldstyle\the\eqcount}}
                \eqno({\oldstyle C}.{\oldstyle\the\eqcount})\fi
\fi}
\def\eqn{\global\advance\eqcount by 1
\ifnum\the\sectioncount=0
	\eqno({\oldstyle\the\eqcount})
\else
        \ifnum\the\appendixcount=0
                \eqno({\oldstyle\the\sectioncount}.{\oldstyle\the\eqcount})\fi
        \ifnum\the\appendixcount=1
                \eqno({\oldstyle A}.{\oldstyle\the\eqcount})\fi
        \ifnum\the\appendixcount=2
                \eqno({\oldstyle B}.{\oldstyle\the\eqcount})\fi
        \ifnum\the\appendixcount=3
                \eqno({\oldstyle C}.{\oldstyle\the\eqcount})\fi
\fi}
\def\multi{\global\advance\eqcount by 1}
\def\multieqn#1{({\oldstyle\the\sectioncount}.{\oldstyle\the\eqcount}\hbox{#1})}
\def\multiEqn#1#2{\xdef#1{{\oldstyle\the\sectioncount}.{\old\the\eqcount}#2}
        ({\oldstyle\the\sectioncount}.{\oldstyle\the\eqcount}\hbox{#2})}
\def\multiEqnAll#1{\xdef#1{{\oldstyle\the\sectioncount}.{\old\the\eqcount}}}
\newcount\tablecount
\tablecount=0
\def\Table#1#2#3{\global\advance\tablecount by 1
\immediate\write\intrefwrite{\def\noexpand#1{{\noexpand\oldsize\the\tablecount}}}
       \vtop{\vskip2\parskip
       \centerline{#2}
       \vskip5\parskip
       \centerline{\it Table \the\tablecount: #3}
       \vskip2\parskip}}
\newcount\figurecount
\figurecount=0
\def\Figure#1#2#3{\global\advance\figurecount by 1
\immediate\write\intrefwrite{\def\noexpand#1{{\noexpand\oldsize\the\figurecount}}}
       \vtop{\vskip2\parskip
       \centerline{#2}
       \vskip4\parskip
       \centerline{\it Figure \the\figurecount: #3}
       \vskip3\parskip}}
\newtoks\url
\def\Href#1#2{\catcode`\#=12\url={#1}\catcode`\#=\active#2}
\def\href#1#2{{#2}}

\parskip=3.5pt plus .3pt minus .3pt
\baselineskip=14pt plus .1pt minus .05pt
\lineskip=.5pt plus .05pt minus .05pt
\lineskiplimit=.5pt
\abovedisplayskip=18pt plus 4pt minus 2pt
\belowdisplayskip=\abovedisplayskip
\hsize=14cm
\vsize=19cm
\hoffset=1.5cm
\voffset=1.8cm
\frenchspacing
\footline={}
\raggedbottom

\newskip\origparindent
\origparindent=\parindent

\def\ss{\scriptstyle}

\def\*{\partial}
\def\punkt{\,\,.}
\def\komma{\,\,,}

\def\={\!=\!}
\def\small#1{{\hbox{$#1$}}}

\def\fraction#1{\small{1\over#1}}
\def\fr{\fraction}
\def\Fraction#1#2{\small{#1\over#2}}
\def\Fr{\Fraction}
\def\tr{\hbox{\rm tr}}
\def\eg{{\it e.g.}}

\def\ie{{\it i.e.}}
\def\etal{{\it et al.}}

\def\a{\alpha}
\def\b{\beta}
\def\e{\varepsilon}
\def\g{\gamma}

\def\G{\Gamma}

\def\id{1\hskip-3.5pt 1}

\def\Dslash{D\hskip-6.5pt/\hskip1.5pt}
\def\dslash{\*\hskip-5.5pt/\hskip.5pt}

\def\II{I\hskip-.8pt I}

\def\RR{{\Bbb R}}




\def\textfrac#1#2{\raise .45ex\hbox{\the\scriptfont0 #1}\nobreak\hskip-1pt/\hskip-1pt\hbox{\the\scriptfont0 #2}}

\def\LL{{\cal L}}
\def\leftbr{[\![}
\def\rightbr{]\!]}


\def\frac{\Fr}

\def\mathbb{\Bbb}

\def\NN{{\Bbb N}}



\def\LL{{\cal L}}
\def\leftbr{[\![}
\def\rightbr{]\!]}

\def\LL{{\cal L}}
\def\leftbr{[\![}
\def\rightbr{]\!]}

\def\ms{{\mathstrut}}

\def\LL{{\cal L}}

\def\leftbr{[\![}
\def\rightbr{]\!]}




\catcode`@=11
\def\openupnormal{\afterassignment\@penupnormal\dimen@=}
\def\@penupnormal{\advance\normallineskip\dimen@
  \advance\normalbaselineskip\dimen@
  \advance\normallineskiplimit\dimen@}
\catcode`@=12

\def\EqMatrix{\let\quad\enspace\openupnormal6pt\matrix}

%
\line{
\epsfysize=18mm
\epsffile{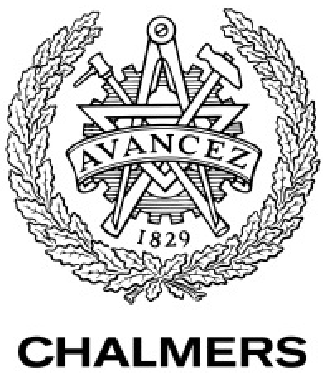}
\hfill}
\vskip-16mm

\line{\hfill}
\line{\hfill Gothenburg preprint}
\line{\hfill March, {\old2016}}
\line{\hrulefill}


\headtext={Cederwall: 
``Double supergeometry''}

\vfill
\vskip.5cm

\centerline{\sixteenhelvbold
Double supergeometry}

%

\vfill

\centerline{\twelvehelvbold Martin Cederwall}

\vfill
\vskip-1cm

\centerline{\it Division for Theoretical Physics}
\centerline{\it Department of Physics}
\centerline{\it Chalmers University of Technology}
\centerline{\it SE 412 96 Gothenburg, Sweden}

\vfill

{\narrower\noindent \underbar{Abstract:}
A geometry of superspace corresponding to double field
theory is developed, with type \II\ supergravity in $D=10$ as the main example.
The formalism is based on an orthosymplectic extension $OSp(d,d|2s)$ of
the continuous T-duality group. Covariance under generalised
super-diffeomorphisms is manifest.
Ordinary superspace is obtained as a solution of the orthosymplectic
section condition.
A systematic study of curved superspace
Bianchi identities is performed,
and a relation to a double pure spinor
superfield cohomology is established. 
A Ramond-Ramond superfield is
constructed as an infinite-dimensional orthosymplectic spinor. Such
objects in minimal orbits under the $OSp$ supergroup (``pure
spinors'') define super-sections.
\smallskip}
\vfill

\font\xxtt=cmtt6

\vtop{\baselineskip=.6\baselineskip\xxtt
\line{\hrulefill}
\catcode`\@=11
\line{email: martin.cederwall@chalmers.se\hfill}
\catcode`\@=\active
}

\eject



\def\textfrac#1#2{\raise .45ex\hbox{\the\scriptfont0 #1}\nobreak\hskip-1pt/\hskip-1pt\hbox{\the\scriptfont0 #2}}

\def\LL{{\cal L}}
\def\leftbr{[\![}
\def\rightbr{]\!]}


\def\frac{\Fr}

\def\mathbb{\Bbb}

\def\Str{\hbox{Str}}

\newskip\scrskip
\scrskip=-.6pt plus 0pt minus .1pt

\def\sM{{\hskip\scrskip{\scr M}\hskip\scrskip}}
\def\sN{{\hskip\scrskip{\scr N}\hskip\scrskip}}
\def\sP{{\hskip\scrskip{\scr P}\hskip\scrskip}}
\def\sQ{{\hskip\scrskip{\scr Q}\hskip\scrskip}}
\def\sR{{\hskip\scrskip{\scr R}\hskip\scrskip}}

\def\sA{{\hskip\scrskip{\scr A}\hskip\scrskip}}
\def\sB{{\hskip\scrskip{\scr B}\hskip\scrskip}}
\def\sC{{\hskip\scrskip{\scr C}\hskip\scrskip}}

\def\mm{{\dot m}}
\def\nn{{\dot n}}
\def\pp{{\dot p}}
\def\qq{{\dot q}}
\def\rr{{\dot r}}

\def\aa{{\dot a}}
\def\bb{{\dot b}}
\def\cc{{\dot c}}
\def\dd{{\dot d}}


\newwrite\intrefwrite
\immediate\openout\intrefwrite=\jobname.intref

\newwrite\contentswrite

\newdimen\sublength
\sublength=\hsize 
\advance\sublength by -\parindent

\newdimen\contentlength
\contentlength=\sublength

\advance\sublength by -\parindent

\def\section#1#2{\global\eqcount=0
	\global\subsectioncount=0
        \global\advance\sectioncount by 1
\ifnum\the\sectioncount=1\immediate\openout\contentswrite=\jobname.contents\fi
\immediate\write\contentswrite{\item{\the\sectioncount.}\hbox to\contentlength{#2\dotfill\the\pageno}}
	\ifnum\sectioncount>1
	        \vskip2\baselineskip
	\fi
\immediate\write\intrefwrite{\def\noexpand#1{{\noexpand\oldsize\the\sectioncount}}}\noindent{\twelvecp\the\sectioncount. #2}\par\nobreak
       \vskip.5\baselineskip\noindent}

\def\subsection#1#2{\global\advance\subsectioncount by 1
\immediate\write\contentswrite{\itemitem{\the\sectioncount.\the\subsectioncount.}\hbox
to\sublength{#2\dotfill\the\pageno}}
\immediate\write\intrefwrite{\def\noexpand#1{{\noexpand\oldsize\the\sectioncount}.{\noexpand\oldsize\the\subsectioncount}}}\vskip.75\baselineskip\noindent\line{\tencp\the\sectioncount.\the\subsectioncount. #2\hfill}\nobreak\vskip.4\baselineskip\nobreak\noindent}

\def\immediatesubsection#1#2{\global\advance\subsectioncount by 1
\immediate\write\contentswrite{\itemitem{\the\sectioncount.\the\subsectioncount.}\hbox
to\sublength{#2\dotfill\the\pageno}}
\immediate\write\intrefwrite{\def\noexpand#1{{\noexpand\oldsize\the\sectioncount}.{\noexpand\oldsize\the\subsectioncount}}}
\vskip-\baselineskip\noindent
\line{\tencp\the\sectioncount.\the\subsectioncount. #2\hfill}
	\vskip.5\baselineskip\noindent}

\def\contentsout{\catcode`\@=11
        \vskip2\baselineskip
        {\noindent\twelvecp Contents}\hfill\vskip\baselineskip
        \input\jobname.contents
        \catcode`\@=\active\rm
\vskip3\baselineskip
}

\def\refout{\eightpoint\catcode`\@=11
        \immediate\write\contentswrite{\item{}\hbox to\contentlength{References\dotfill\the\pageno}}
        \xrm\immediate\closeout\refwrite
        \vskip2\baselineskip
        {\noindent\twelvecp References}\hfill\vskip\baselineskip
        \baselineskip=.75\baselineskip
        \input\jobname.refs
        \baselineskip=4\baselineskip \divide\baselineskip by 3
        \catcode`\@=\active\rm}


\ref\HatsudaKamimuraSiegelI{M. Hatsuda, K. Kamimura and W. Siegel,
{\xit ``Superspace with manifest T-duality from type II
superstring''}, \jhep{14}{06}{2014}{039} [\arxiv{1403}{3887}].}

\ref\HatsudaKamimuraSiegelII{M. Hatsuda, K. Kamimura and W. Siegel,
{\xit ``Ramond-Ramond gauge fields in superspace with manifest
T-duality''}, \jhep{15}{02}{2015}{134} [\arxiv{1411}{2206}].} 

\ref\Duff{M.J. Duff, {\xit ``Duality rotations in string
theory''}, \NPB{335}{1990}{610}.}

\ref\Tseytlin{A.A.~Tseytlin,
  {\xit ``Duality symmetric closed string theory and interacting
  chiral scalars''}, 
  \NPB{350}{1991}{395}.}

\ref\SiegelI{W.~Siegel,
  {\xit ``Two vierbein formalism for string inspired axionic gravity''},
  \PRD{47}{1993}{5453}
  [\hepth{9302036}].}

\ref\SiegelII{ W.~Siegel,
  {\xit ``Superspace duality in low-energy superstrings''},
  \PRD{48}{1993}{2826}
  [\hepth{9305073}].}

\ref\SiegelIII{W.~Siegel,
  {\xit ``Manifest duality in low-energy superstrings''},
  in Berkeley 1993, Proceedings, Strings '93 353
  [\hepth{9308133}].}

\ref\HullDoubled{C.M. Hull, {\xit ``Doubled geometry and
T-folds''}, \jhep{07}{07}{2007}{080}
[\hepth{0605149}].}

\ref\HullT{C.M. Hull, {\xit ``A geometry for non-geometric string
backgrounds''}, \jhep{05}{10}{2005}{065} [\hepth{0406102}].}

\ref\HullM{C.M. Hull, {\xit ``Generalised geometry for M-theory''},
\jhep{07}{07}{2007}{079} [\hepth{0701203}].}

\ref\HullZwiebachDFT{C. Hull and B. Zwiebach, {\xit ``Double field
theory''}, \jhep{09}{09}{2009}{99} [\arxiv{0904}{4664}].}

\ref\HohmHullZwiebachI{O. Hohm, C.M. Hull and B. Zwiebach, {\xit ``Background
independent action for double field
theory''}, \jhep{10}{07}{2010}{016} [\arxiv{1003}{5027}].}

\ref\HohmHullZwiebachII{O. Hohm, C.M. Hull and B. Zwiebach, {\xit
``Generalized metric formulation of double field theory''},
\jhep{10}{08}{2010}{008} [\arxiv{1006}{4823}].} 

\ref\HohmKwak{O. Hohm and S.K. Kwak, {\xit ``$N=1$ supersymmetric
double field theory''}, \jhep{12}{03}{2012}{080} [\arxiv{1111}{7293}].}

\ref\HohmKwakFrame{O. Hohm and S.K. Kwak, {\xit ``Frame-like geometry
of double field theory''}, \JPA{44}{2011}{085404} [\arxiv{1011}{4101}].}

\ref\JeonLeeParkI{I. Jeon, K. Lee and J.-H. Park, {\xit ``Differential
geometry with a projection: Application to double field theory''},
\jhep{11}{04}{2011}{014} [\arxiv{1011}{1324}].}

\ref\JeonLeeParkII{I. Jeon, K. Lee and J.-H. Park, {\xit ``Stringy
differential geometry, beyond Riemann''}, 
\PRD{84}{2011}{044022} [\arxiv{1105}{6294}].}

\ref\JeonLeeParkIII{I. Jeon, K. Lee and J.-H. Park, {\xit
``Supersymmetric double field theory: stringy reformulation of supergravity''},
\PRD{85}{2012}{081501} [\arxiv{1112}{0069}].}

\ref\HohmZwiebachLarge{O. Hohm and B. Zwiebach, {\xit ``Large gauge
transformations in double field theory''}, \jhep{13}{02}{2013}{075}
[\arxiv{1207}{4198}].} 

\ref\Park{J.-H.~Park,
  {\xit ``Comments on double field theory and diffeomorphisms''},
  \jhep{13}{06}{2013}{098}
  [\arxiv{1304}{5946}].}

\ref\BermanCederwallPerry{D.S. Berman, M. Cederwall and M.J. Perry,
{\xit ``Global aspects of double geometry''}, 
\jhep{14}{09}{2014}{66} [\arxiv{1401}{1311}].}

\ref\CederwallGeometryBehind{M. Cederwall, {\xit ``The geometry behind
double geometry''}, 
\jhep{14}{09}{2014}{70} [\arxiv{1402}{2513}].}

\ref\PachecoWaldram{P.P. Pacheco and D. Waldram, {\xit ``M-theory,
exceptional generalised geometry and superpotentials''},
\jhep{08}{09}{2008}{123} [\arxiv{0804}{1362}].}

\ref\Hillmann{C. Hillmann, {\xit ``Generalized $E_{7(7)}$ coset
dynamics and $D=11$ supergravity''}, \jhep{09}{03}{2009}{135}
[\arxiv{0901}{1581}].}

\ref\BermanPerryGen{D.S. Berman and M.J. Perry, {\xit ``Generalised
geometry and M-theory''}, \jhep{11}{06}{2011}{074} [\arxiv{1008}{1763}].}    

\ref\BermanGodazgarPerry{D.S. Berman, H. Godazgar and M.J. Perry,
{\xit ``SO(5,5) duality in M-theory and generalized geometry''},
\PLB{700}{2011}{65} [\arxiv{1103}{5733}].} 

\ref\BermanMusaevPerry{D.S. Berman, E.T. Musaev and M.J. Perry,
{\xit ``Boundary terms in generalized geometry and doubled field theory''},
\PLB{706}{2011}{228} [\arxiv{1110}{97}].} 

\ref\BermanGodazgarGodazgarPerry{D.S. Berman, H. Godazgar, M. Godazgar  
and M.J. Perry,
{\xit ``The local symmetries of M-theory and their formulation in
generalised geometry''}, \jhep{12}{01}{2012}{012}
[\arxiv{1110}{3930}].} 

\ref\BermanGodazgarPerryWest{D.S. Berman, H. Godazgar, M.J. Perry and
P. West,
{\xit ``Duality invariant actions and generalised geometry''}, 
\jhep{12}{02}{2012}{108} [\arxiv{1111}{0459}].} 

\ref\CoimbraStricklandWaldram{A. Coimbra, C. Strickland-Constable and
D. Waldram, {\xit ``$E_{d(d)}\times\hbox{\eightbbb R}^+$ generalised geometry,
connections and M theory'' }, \jhep{14}{02}{2014}{054} [\arxiv{1112}{3989}].} 

\ref\CoimbraStricklandWaldramII{A. Coimbra, C. Strickland-Constable and
D. Waldram, {\xit ``Supergravity as generalised geometry II:
$E_{d(d)}\times\hbox{\eightbbb R}^+$ and M theory''}, 
\jhep{14}{03}{2014}{019} [\arxiv{1212}{1586}].}  

\ref\JeonLeeParkSuh{I. Jeon, K. Lee, J.-H. Park and Y. Suh, {\xit
``Stringy unification of Type IIA and IIB supergravities under N=2
D=10 supersymmetric double field theory''}, \PLB{723}{2013}{245}
[\arxiv{1210}{5048}].} 

\ref\JeonLeeParkRR{I. Jeon, K. Lee and J.-H. Park, {\xit
``Ramond--Ramond cohomology and O(D,D) T-duality''},
\jhep{12}{09}{2012}{079} [\arxiv{1206}{3478}].} 

\ref\BermanCederwallKleinschmidtThompson{D.S. Berman, M. Cederwall,
A. Kleinschmidt and D.C. Thompson, {\xit ``The gauge structure of
generalised diffeomorphisms''}, \jhep{13}{01}{2013}{64} [\arxiv{1208}{5884}].}

\ref\ParkSuh{J.-H. Park and Y. Suh, {\xit ``U-geometry: SL(5)''},
\jhep{14}{06}{2014}{102} [\arxiv{1302}{1652}].} 

\ref\CederwallI{M.~Cederwall, J.~Edlund and A.~Karlsson,
  {\xit ``Exceptional geometry and tensor fields''},
  \jhep{13}{07}{2013}{028}
  [\arxiv{1302}{6736}].}

\ref\CederwallII{ M.~Cederwall,
  {\xit ``Non-gravitational exceptional supermultiplets''},
  \jhep{13}{07}{2013}{025}
  [\arxiv{1302}{6737}].}

\ref\SambtlebenHohmI{O.~Hohm and H.~Samtleben,
  {\xit ``Exceptional field theory I: $E_{6(6)}$ covariant form of
  M-theory and type IIB''}, 
  \PRD{89}{2014}{066016} [\arxiv{1312}{0614}].}

\ref\SambtlebenHohmII{O.~Hohm and H.~Samtleben,
  {\xit ``Exceptional field theory II: $E_{7(7)}$''},
  \PRD{89}{2014}{066016} [\arxiv{1312}{4542}].}

\ref\HohmSamtlebenIII{O. Hohm and H. Samtleben, {\xit ``Exceptional field
theory III: $E_{8(8)}$''}, \PRD{90}{2014}{066002} [\arxiv{1406}{3348}].}

\ref\KachruNew{S. Kachru, M.B. Schulz, P.K. Tripathy and S.P. Trivedi,
{\xit ``New supersymmetric string compactifications''}, 
\jhep{03}{03}{2003}{061} [\hepth{0211182}].}

\ref\Condeescu{C. Condeescu, I. Florakis, C. Kounnas and D. L\"ust, 
{\xit ``Gauged supergravities and non-geometric $Q$/$R$-fluxes from
asymmetric orbifold CFT's''}, 
\jhep{13}{10}{2013}{057} [\arxiv{1307}{0999}].}

\ref\CederwallUfoldBranes{M. Cederwall, {\xit ``M-branes on U-folds''},
in proceedings of 7th International Workshop ``Supersymmetries and
Quantum Symmetries'' Dubna, 2007 [\arxiv{0712}{4287}].}

\ref\HasslerLust{F. Hassler and D. L\"ust, {\xit ``Consistent
compactification of double field theory on non-geometric flux
backgrounds''}, \jhep{14}{05}{2014}{085} [\arxiv{1401}{5068}].}

\ref\CederwallDuality{M. Cederwall, {\xit ``T-duality and
non-geometric solutions from double geometry''}, \FP{62}{2014}{942}
[\arxiv{1409}{4463}].} 

\ref\CederwallRosabal{M. Cederwall and J.A. Rosabal, ``$E_8$
geometry'', \jhep{15}{07}{2015}{007}, [\arxiv{1504}{04843}].}

\ref\HohmKwakZwiebachI{O. Hohm, S.K. Kwak and B. Zwiebach, {\xit
``Unification of type II strings and T-duality''}, \PRL{107}{2011}{171603}
[\arxiv{1106}{5452}].}  

\ref\HohmZwiebachGeometry{O. Hohm and B. Zwiebach, {\xit ``Towards an
invariant geometry of double field theory''}, \arxiv{1212}{1736}.} 

\ref\CGNT{M. Cederwall, U. Gran, B.E.W. Nilsson and D. Tsimpis,
{\xit ``Supersymmetric corrections to eleven-dimen\-sional supergravity''},
\jhep{05}{05}{2005}{052} [\hepth{0409107}].}

\ref\CartanSpinors{E. Cartan, {\xit ``Le\hskip.5pt,\hskip-3.5pt cons sur
la th\'eorie des spineurs''} (Hermann, Paris, 1937).}

\ref\AsakawaSasaWatamura{T. Asakawa, S. Sasa and S. Watamura, {\xit
``D-branes in generalized geometry and Dirac--Born--Infeld action''},
\jhep{12}{10}{2012}{064} [\arxiv{1206}{6964}].} 

\ref\BermanCederwallMalek{D.S. Berman, M. Cederwall and E. Malek,
{\xit work in progress}.}

\ref\DBranesII{M. Cederwall, A. von Gussich, B.E.W. Nilsson, P. Sundell
 and A. Westerberg,
{\xit ``The Dirichlet super-p-branes in ten-dimensional type IIA and IIB 
supergravity''},
\NPB{490}{1997}{179} [\hepth{9611159}].}

\ref\AzcarragaTownsend{J.A. de Azc\'arraga and P.K. Townsend, {\xit ``Superspace geometry and classification of supersymmetric extended objects''}, \PRL{62}{1989}{2579}.}

\ref\BerkovitsI{N. Berkovits, 
{\xit ``Super-Poincar\'e covariant quantization of the superstring''}, 
\jhep{00}{04}{2000}{018} [\hepth{0001035}].}

\ref\BerkovitsParticle{N. Berkovits, {\xit ``Covariant quantization of
the superparticle using pure spinors''}, \jhep{01}{09}{2001}{016}
[\hepth{0105050}].}

\ref\CederwallNilssonTsimpisI{M. Cederwall, B.E.W. Nilsson and D. Tsimpis,
{\xit ``The structure of maximally supersymmetric super-Yang--Mills
theory --- constraining higher order corrections''},
\jhep{01}{06}{2001}{034} 
[\hepth{0102009}].}

\ref\CederwallNilssonTsimpisII{M. Cederwall, B.E.W. Nilsson and D. Tsimpis,
{\xit ``D=10 super-Yang--Mills at $\ss O(\a'^2)$''},
\JHEP{01}{07}{2001}{042} [\hepth{0104236}].}

\ref\SpinorialCohomology{M. Cederwall, B.E.W. Nilsson and D. Tsimpis, 
{\xit ``Spinorial cohomology and maximally supersymmetric theories''},
\jhep{02}{02}{2002}{009} [\hepth{0110069}];
M. Cederwall, {\xit ``Superspace methods in string theory,
supergravity and gauge theory''}, Lectures at the XXXVII Winter School
in Theoretical Physics ``New Developments in Fundamental Interactions
Theories'',  Karpacz, Poland,  Feb. 6-15, 2001, \hepth{0105176}.}

\ref\CederwallBLG{M. Cederwall, {\xit ``N=8 superfield formulation of
the Bagger--Lambert--Gustavsson model''}, \jhep{08}{09}{2008}{116}
[\arxiv{0808}{3242}].}

\ref\CederwallABJM{M. Cederwall, {\xit ``Superfield actions for N=8 
and N=6 conformal theories in three dimensions''},
\jhep{08}{10}{2008}{70}
[\arxiv{0809}{0318}].}

\ref\PureSGI{M. Cederwall, {\xit ``Towards a manifestly supersymmetric
    action for D=11 supergravity''}, \jhep{10}{01}{2010}{117}
    [\arxiv{0912}{1814}].}  

\ref\PureSGII{M. Cederwall, 
{\xit ``D=11 supergravity with manifest supersymmetry''},
    \MPLA{25}{2010}{3201} [\arxiv{1001}{0112}].}

\ref\PureSpinorOverview{M. Cederwall, {\xit ``Pure spinor superfields
--- an overview''}, Springer Proc. Phys. {\xbf153} ({\xrm2013}) {\xrm61} 
[\arxiv{1307}{1762}].}

\ref\KacSuperalgebras{V.G. Kac, {\xit ``Classification of simple Lie
superalgebras''}, Funktsional. Anal. i Prilozhen. {\xbold9}
({\xold1975}) {\xold91}.}

\ref\CederwallExceptionalTwistors{M. Cederwall, {\xit ``Twistors and
supertwistors for exceptional field
theory''}, \jhep{15}{12}{2015}{123} [\arxiv{1510}{02298}].}

\ref\MaDBrane{C.-T. Ma, {\xit ``Gauge transformation of double field
theory for open string''}, \PRD{92}{2015}{066004} [\arxiv{1411}{0287}].}

\ref\BandosStringsSuperspace{I Bandos, {\xit ``Strings in doubled
superspace''}, \PLB{751}{2015}{402} [\arxiv{1507}{07779}].}

\ref\BandosESeven{I Bandos, {\xit ``On section conditions of
$E_{7(+7)}$ exceptional field theory and superparticle in N=8 central
charge superspace''}, \jhep{16}{01}{2016}{132} [\arxiv{1512}{02287}].}


\contentsout

\section\IntroductionSection{Introduction}There is by now a
significant bulk of
work on double geometry
[\SiegelII\skipref\HatsudaKamimuraSiegelI\skipref\HatsudaKamimuraSiegelII\skipref\Duff\skipref\Tseytlin\skipref\SiegelI\skipref\SiegelIII\skipref\HullT\skipref\HullDoubled\skipref\HullZwiebachDFT\skipref\HohmHullZwiebachI\skipref\HohmHullZwiebachII\skipref\HohmKwakFrame\skipref\HohmKwak\skipref\JeonLeeParkI\skipref\JeonLeeParkII\skipref\JeonLeeParkIII\skipref\HohmZwiebachGeometry\skipref\HohmKwakZwiebachI\skipref\JeonLeeParkSuh\skipref\JeonLeeParkRR\skipref\HohmZwiebachLarge\skipref\Park\skipref\BermanCederwallPerry\skipref\CederwallGeometryBehind-\CederwallDuality]
and exceptional geometry
[\HullM\skipref\PachecoWaldram\skipref\Hillmann\skipref\BermanPerryGen\skipref\BermanGodazgarPerry\skipref\BermanGodazgarGodazgarPerry\skipref\BermanGodazgarPerryWest\skipref\CoimbraStricklandWaldram\skipref\CoimbraStricklandWaldramII\skipref\BermanCederwallKleinschmidtThompson\skipref\ParkSuh\skipref\CederwallI\skipref\CederwallII\skipref\CederwallUfoldBranes\skipref\SambtlebenHohmI\skipref\SambtlebenHohmII\skipref\HohmSamtlebenIII\skipref\CederwallRosabal-\CederwallExceptionalTwistors]. Although
some of the work concerns supersymmetry, little
attention has been given to supergeometric formulations. In double
geometry there are a few papers by Siegel and by Hatsuda \etal\
[\SiegelII,\HatsudaKamimuraSiegelI,\HatsudaKamimuraSiegelII], and in
the exceptional setting nothing, to the best of our knowledge.
A few papers
[\CederwallExceptionalTwistors,\BandosStringsSuperspace,\BandosESeven]
deal with particles and strings in flat superspace.
The purpose of the present paper is to investigate double
supergeometry, starting from diffeomorphisms on a double
superspace. The generalised diffeomorphisms and the corresponding
local supersymmetry will thus be manifest in the formalism. 

There are several obvious motivations for trying to achieve a
formulation where these symmetries are manifest. The foremost may be
the belief that a formulation with as much symmetry as possible
manifested should be more elegant and perhaps simpler, but there are
also more practical issues like the construction of terms in
supergravity effective
actions restricted both by maximal supersymmetry and by duality. 

Our approach closely parallels the geometric formulation of double
geometry [\HohmZwiebachGeometry], and provides a natural
supersymmetric counterpart. As will be argued, a doubled
superspace indeed has twice the number of coordinates, both bosonic and
fermionic, as an ordinary superspace. The latter will
arise as a solution of the supersymmetric section condition. The
r\^ole of the group $O(d,d)$ in double geometry is subsumed by an
orthosymplectic group $OSp(d,d|2s)$.
Unlike previous work on double supergeometry
[\SiegelII,\HatsudaKamimuraSiegelI,\HatsudaKamimuraSiegelII], our
formalism does not 
use any input from string theory in terms of world-sheet algebras. Our
superspace is also considerably smaller, and uses a locally realised
symmetry (some real form of) $Spin(d)\times Spin(d)$, which ties more
directly both to double geometry and to ordinary superspace.

First, a review of double geometry will be given in
section \BackgroundSection. It will 
provide the necessary tools to extend to double supergeometry in
section \DoubleSuperGeometrySection. 
There, it will be argued that the Bianchi identities take the
theory on shell (in the case of maximal supersymmetry). Support is
obtained from the interpretation in terms of a double pure spinor
superfield in section \FieldsSection. Section \RRSection\ 
contains a double superspace
formulation of the Ramond-Ramond fields. The structures examined there
provide some clues to what will happen in a exceptional setting; this
is discussed in the conclusions in section \ConclusionSection, where a further
comparison to earlier work also is made.

\section\BackgroundSection{Background --- double geometry}The
geometric formulation of double field theory [\HohmZwiebachGeometry]
(see also refs. [\SiegelIII,\HohmKwakFrame,\JeonLeeParkII])
is well known.
It will be briefly recapitulated here, since most of the calculations in double
supergeometry that will be performed in section \DoubleSuperGeometrySection\
closely parallel the ones in the bosonic situation. 
In fact, much of the information will be obtained fairly
directly by replacing (anti-)symmetrisations with graded versions.

Let the coordinate basis tangent vectors of doubled space carry an
index $\dot m$. The coordinates $X^{\dot m}$ are the bosonic part of
the superspace coordinates of the
following section\foot{The standard notation is to use an index $M$, but
this will be reserved for another set of superspace coordinates. The
somewhat awkward index convention is consequence of the need of a large
number of alphabets later in the paper.}.

The tangent space of doubled space, which is identical to the
generalised/doubled tangent space of ordinary space in generalised
geometry, is equipped with an $O(d,d)$ structure defined by an
$O(d,d)$-invariant metric $\eta_{\mm\nn}$. In a suitable basis, where 
$dX^\mm=(dx^m,d\tilde x_m)$ it takes the form
$$
\eta_{\mm,\nn}=\left(\matrix{0&\delta_m{}^n\cr\delta^m{}_n&0}\right)\punkt\eqn
$$

The section condition, the solutions of which locally reduces the
dependence of fields on coordinates to that of an ordinary
$d$-dimensional space, reads
$$
\eta^{\mm\nn}\*_\mm\otimes\*_\nn=0\punkt\Eqn\BosonicSectionCondition
$$
The meaning of ``$\otimes$'' is that the two derivatives can act on
any fields. This means that one looks for a linear subspace of momenta,
where all vectors mutually satisfy (\BosonicSectionCondition). Modulo
the choice of basis for $O(d,d)$, the solution is given by
${\*\over\*\tilde x_m}=0$, so fields locally depend only on $x^m$.

Generalised diffeomorphisms with a parameter $\xi^\mm$ transform
(co)vectors according with the generalised Lie derivative (the
Dorfman bracket)
$$
\LL_\xi V_\mm=\xi V_\mm+(a-a^T)_\mm{}^\nn V_\nn\komma\eqn
$$
where $\xi=\xi^\mm\*_\mm$ and $a_\mm{}^\nn=\*_\mm\xi^\nn$, and where
$a^T=\eta a^t\eta^{-1}$. The transformations commute to
$$
[\LL_\xi,\LL_\eta]=\LL_{\leftbr\xi,\eta\rightbr}\komma\eqn
$$
where
$$
\leftbr\xi,\eta\rightbr=\fr2(\LL_\xi\eta-\LL_\eta\xi)\punkt\eqn
$$
is the Courant bracket.
The Courant bracket does not satisfy a Jacobi identity, but the
violation is a parameter of the form
$\zeta^\mm=\eta^{\mm\nn}\*_\nn\lambda$ with $\LL_\zeta=0$,
representing the singlet reducibility of the gauge transformation
(directly inherited from the second order, 
gauge for gauge, transformation for the
$B$ field).

The generalised geometric fields, the metric and the $B$-field, are
encoded in a generalised metric or a generalised vielbein. Here, as is
normal for superspace,
the latter is chosen. The vielbein $E_\mm{}^\aa$ is a group element of
$O(d,d)$. It is demanded to be covariantly constant when transported
by a covariant derivative with generalised affine and spin
connections,
$$
D_\mm E_\nn{}^\aa=\*_\mm E_\nn{}^\aa+\Gamma_{\mm\nn}{}^\pp E_\pp{}^\aa
-E_\nn{}^\bb\Omega_{\mm\bb}{}^\aa=0\punkt\Eqn\DFTCompatibility
$$
Covariance of the covariant derivative dictates that $\Gamma$
transforms inhomogeneously under generalised diffeomorphisms:
$$
\delta_\xi\Gamma_{\mm\nn}{}^\pp=\LL_\xi\Gamma_{\mm\nn}{}^\pp
-\*_\mm(\*_\nn\xi^\pp-\*^\pp\xi_\nn)\punkt\eqn
$$
This implies that the totally antisymmetric part
$\Gamma_{[\mm\nn\pp]}$ transforms covariantly. It is defined as
torsion\foot{This normalisation, which differs from the one in
ref. [\HohmZwiebachGeometry], is conventional for later simplification, and
has essentially to do with the retraining of the torsion 
from 2-form to ``3-form''.},
$$
T_{\mm\nn\pp}=-\Fr32\Gamma_{[\mm\nn\pp]}\punkt\Eqn\TorsionNormalisation
$$

Next, curvature is constructed. This can be done as in
ref. [\HohmZwiebachGeometry], by considering combinations of
$\*\Gamma$ and $\Gamma^2$ for which the inhomogeneous transformations
cancel. Another possibility is to use the compatibility equation
(\DFTCompatibility). Then one will also get information about how the
curvature is expressed in terms of the spin connection. The idea is
the standard one: by taking derivatives of the compatibility equation,
one searches for an equality between two expressions, of which one is
manifestly covariant under transformations in the locally realised
subalgebra $so(d)\oplus so(d)$, and the other is a tensor. Then one
concludes that the two expressions, which are the expressions for
curvature in terms of affine connection and spin connection,
respectively, are equal and share both covariance properties.

Taking one derivative on eq. (\DFTCompatibility), antisymmetrising,
and replacing the derivatives on the vielbein
leads to
$$
(\*_{[\mm}\Gamma_{\nn]}+\Gamma_{[\mm}\Gamma_{\nn]})_\pp{}^\qq
E_\qq{}^\aa
-E_\pp{}^\bb(\*_{[\mm}\Omega_{\nn]}+\Omega_{[\mm}\Omega_{\nn]})_\bb{}^\aa=0\punkt
\Eqn\BadCurvature
$$
This is precisely the step taken in ordinary geometry to derive the
Riemann tensor. What fails in double geometry is that
$\*_{[\mm}\Omega_{\nn]}$ is not a tensor, since
$\Gamma_{[\mm\nn]}{}^\pp$ is not torsion. 
Eq. (\BadCurvature) certainly holds, but can not be used to extract
any useful (covariant) information.
The
derivative on $\Omega$ can be covariantised, but one needs to 
compensate with a $\Gamma\Omega$
term. 

In
refs. [\SiegelII,\HohmZwiebachGeometry], a 4-index curvature was constructed by
symmetrisation in the pairs of indices. Explicitly,
$$
\eqalign{
&(\*_{[\mm}\Gamma_{\nn]}+\Gamma_{[\mm}\Gamma_{\nn]})_{\pp\qq}
        -\fr4\Gamma_{\rr\mm\nn}\Gamma^\rr{}_{\pp\qq}\cr
&-E_\pp{}^\aa E_\qq{}^\bb
         (D^{(\Gamma)}_{[\mm}\Omega_{\nn]}+\Omega_{[\mm}\Omega_{\nn]})_{\aa\bb}\cr
&\qquad+E_\pp{}^\aa E_\qq{}^\bb\Gamma_{[\mm\nn]}{}^\rr\Omega_{\rr\aa\bb}
              +\fr4\Gamma_{\rr\mm\nn}\Gamma^\rr{}_{\pp\qq}\cr
&+(\mm\nn\leftrightarrow\pp\qq)=0\punkt\cr
}\Eqn\CurvatureIntermediary
$$
The combination of connections on the third line turns out to be a
tensor under generalised diffeomorphisms. If the vielbeins, for simplicity,
are suppressed, the compatibility equation tells us that
$\Gamma_\mm-\Omega_\mm$ carries a derivative with index
$\mm$. Therefore, the section condition can be used to obtain
$$
\eqalign{
0&=\fr2(-\Gamma_\rr+\Omega_\rr)_{\mm\nn}(-\Gamma^\rr+\Omega^\rr)_{\pp\qq}\cr
&=\fr4\Gamma_{\rr\mm\nn}\Gamma^\rr{}_{\pp\qq}
+(T_{\mm\nn\rr}+\Gamma_{[\mm\nn]\rr})\Omega_{\rr\pp\qq}
+\fr4\Omega_{\rr\mm\nn}\Omega^\rr{}_{\pp\qq}+(\mm\nn\leftrightarrow\pp\qq)
\punkt\cr
}\eqn
$$ 
The only non-tensorial terms (under generalised diffeomorphisms) are
the ones containing $\Gamma$, and match the ones on the third line of
eq. (\CurvatureIntermediary). On the other hand, the first line in
eq. (\CurvatureIntermediary) is manifestly invariant under the local
subgroup. This means that there is a curvature tensor:
$$\eqalign{
R_{\mm\nn\pp\qq}&=(\*_{[\mm}\Gamma_{\nn]}+\Gamma_{[\mm}\Gamma_{\nn]})_{\pp\qq}
        -\fr4\Gamma_{\rr\mm\nn}\Gamma^\rr{}_{\pp\qq}
        +(\mm\nn\leftrightarrow\pp\qq)\cr
&=E_\pp{}^\aa E_\qq{}^\bb
         (D^{(\Gamma)}_{[\mm}\Omega^\ms_{\nn]}+\Omega_{[\mm}\Omega_{\nn]}
         +T_{\mm\nn}{}^\rr\Omega_\rr){}_{\aa\bb}\cr
&\qquad+\fr4E_\mm{}^\aa E_\nn{}^\bb E_\pp{}^\cc E_\qq{}^\dd
                 \Omega_{\rr\aa\bb}\Omega^\rr{}_{\cc\dd}
+(\mm\nn\leftrightarrow\pp\qq)\punkt\cr
}\Eqn\DFTCurvature
$$
By construction, the curvature is antisymmetric in $[\mm\nn]$ and in
$[\pp\qq]$, and symmetric under interchange of the pairs of
indices. The completely antisymmetric part figures in the torsion
Bianchi identity, see below.

The second form of $R$ is maybe not very useful compared to the first
one. However, it conveys one very important piece of information,
which is not manifest from its expression in the affine connection,
namely, that it only gets contributions from terms where at least one
of the two pairs takes values in $so(d)\oplus so(d)$. The
corresponding property of the curvature on superspace will be relied
upon when the superspace Bianchi identities are investigated.

In addition to the generalised metric, a generalised dilaton $\Phi$
is introduced. It transformation is given by
$$
\delta_\xi e^{-2\Phi}=\*_\mm(\xi^\mm e^{-2\Phi})\komma\eqn
$$
which means that $e^{-2\Phi}$ is a density with weight $w=1$. The
action of the covariant derivative $D_\mm$ on a density contains an
extra term $w\Gamma_{\nn\mm}{}^\nn\psi$
for any tensor density $\psi$. Clearly, $e^{2\Phi}D_\mm e^{-2\Phi}$ is a
vector, and it can be seen as an additional part of torsion, $T_\mm$.
It can be used (with knowledge of the dilaton field) to determine a
vector part of the connection,
$$
\G_{\nn\mm}{}^\nn=2\*_\mm \Phi+T_\mm\punkt\eqn
$$

\Yboxdim5pt
\Yvcentermath1

Unlike in ordinary geometry, the compatibility of the vielbein
together with specification of the torsion is not sufficient to
determine the connections completely. There are certain $so(d)\oplus
so(d)$-modules, where torsion is absent, and eq. (\DFTCompatibility)
only gives information about $\Gamma-\Omega$. These modules are the
irreducible hooks, \lower1.5pt\hbox{\yng(2,1)}, under the two
$so(d)$'s.
Any physical relations, such as the equations of motion for the
generalised vielbein, must avoid using the undefined connections. This
turns out to be the case when they are formally derived by variation
of the pseudo-action
$$
S=\int d^{2d}X e^{-2\Phi}R\komma\eqn
$$
$R$ being the scalar curvature obtained from the curvature in
eq. (\DFTCurvature). See ref. [\HohmZwiebachGeometry] for
details. 

When later superspace geometry is investigated, torsion will be non-vanishing,
and we need the torsion Bianchi identity.
A direct calculation leads to the identity\foot{The purpose of the
normalisation of the torsion tensor in 
eq. (\TorsionNormalisation) was to obtain natural coefficients here
($4={4!\over3!}$, $6={4!\over(2!)^2}$).}
$$
4D_{[\mm}T_{\nn\pp\qq]}+6T_{[\mm\nn}{}^\rr
T_{\pp\qq]\rr}=-3R_{[\mm\nn\pp\qq]}
\punkt\eqn
$$

\section\DoubleSuperGeometrySection{Double supergeometry}The vielbein
of double of exceptional geometry is restricted to be a 
group element of the duality group $O(d,d)$ or
$E_{n(n)}\times\RR^+$. If supersymmetry is to be made manifest in
terms of superfields, it seems necessary that superspace,
even in the flat case, has an interpretation as supergeometry.
It does not seem consistent with super-diffeomorphisms to constrain a
bosonic corner of a super-vielbein, or even to consider a bosonic part
of a vector to transform under a restricted subgroup. A direct
solution would entail a super-extension of the duality
group. For double geometry, the T-duality groups allow for natural
extensions in the form of 
supergroups $OSp(d,d|2s)$.

This may na\"\i vely seem to rule out supergeometry based on
exceptional U-duality groups, due to the ``non-existence'' of
exceptional supergroups. We will come back to this issue in the
conclusions, as it turns out that the behaviour of the Ramond-Ramond
fields in double supergeometry points strongly towards
infinite-dimensional supergroups in the exceptional cases.

\subsection\OSpBasicsSubSection{Notation and $OSp$ basics}In 
section \GenSuperDiffSubSection, generalised super-diffeomorphisms
on a $(2d|2s)$-dimensional superspace will be defined. 
The coordinate differentials
form the fundamental module of the orthosymplectic supergroup 
$OSp(d,d|2s)$, and this structure will
be preserved by generalised super-diffeomorphisms, in complete analogy
to the $O(d,d)$ structure in double geometry.

The orthosymplectic group allows an invariant metric
$H_{\sM\sN}$ on
superspace\foot{It seems quite clear that it may be possible to
introduced a arbitrary, possibly curved, superspace metric instead of
$H$, along the lines of
refs. [\CederwallGeometryBehind,\CederwallDuality]. That option is not
explored here, 
but a constant, algebraic, $H$ is used.}. A basis may be chosen so that
$$
H_{\sM\sN}=\left(\matrix{\eta_{\dot m\dot n}&0\cr    
  0&\epsilon_{\dot\mu\dot\nu}\cr}\right)
\komma\eqn
$$
where $\eta$ is the symmetric $O(d,d)$ metric, and $\e$ is the
antisymmetric invariant tensor of $Sp(2s)$.
Thus, tangent indices $\sM=(\dot m;\dot\mu)$ are used in coordinate basis. 
Later, when a generalised super-vielbein is introduced, there will
also be flat indices. The locally realised subgroup (analogous to the
Lorentz group in ordinary superspace) will be a real form of
$Spin(d)\times Spin(d)$, and the corresponding indices are denoted 
$\sA=(\dot a;\dot\alpha)=(a',a'';\alpha',\alpha'')$. Here the dotted
indices are the collective $2d$ and $2s$ flat indices, which are
vectors and spinors, respectively, under the two factors of
$Spin(d)$. Vectors and spinors under the first and second $Spin(d)$ are denoted
with primed and double-primed indices 
(this refers to type \II, which will be our main
interest).

\Figure\IndexFigure{\vtop{
$$
\sarrowlength=18pt 
\matrix{&&\underline{OSp(d,d|2s):}&&\cr
&&&&\cr
&&\sM=(\mm;\dot\mu)&&\cr
&&=(M,\bar M)&&\cr
&\arrow(-1,-1)&&\arrow(1,-1)&\cr
&&&&\underline{Spin(d)\times Spin(d):}\cr
\underline{GL(d|s):}&&&&\cr
&&&&\sA=(\dot a;\dot\alpha)=(\dot a;\alpha,\bar\alpha)\cr
M=(m;\mu)&&&&            =(a',a'';\alpha',\alpha'',\bar\alpha',\bar\alpha'') 
                                                        \cr
&&&&=(A,\bar A)\cr
&\arrow(1,-1)&&\arrow(-1,-1)&\cr
&&\underline{Spin(d):}&&\cr
&&&&\cr
&&A=(a;\alpha)&&
}
$$
}}
{Summary of 
index notation for the various relevant groups
and supergroups.}

Another natural decomposition of $\sM$ is in terms of a 
$GL(d|s)\subset OSp(d,d|2s)$
subgroup. Then $dZ^\sM=(dX^{\dot m};d\Theta^{\dot\mu})
=(dx^m,d\tilde x_m;d\theta^\mu,d\tilde\theta_\mu)$, and the invariant
metric takes the form
$$
H_{\sM\sN}=\left(\matrix{0&\delta_m{}^n&0&0\cr
                \delta^m{}_n&0&0&0\cr
                0&0&0&\delta_\mu{}^\nu\cr
                0&0&-\delta^\mu{}_\nu         
       }\right)\punkt\Eqn\InvMetricGL
$$
This decomposition is relevant for local solutions to the strong
section condition (see below). Occasionally, the indices 
$dz^M=(dx^m,d\theta^\mu)$ and $d\tilde z_M=(d\tilde
x_m,d\tilde\theta_\mu)$ are used. The coordinates $z^M$ span an ordinary
superspace. The corresponding flat indices are written $A=(a,\alpha)$,
a vector and spinor under the diagonal subgroup $Spin(d)$ which is also
a subgroup of $GL(d|s)$.
The index conventions are summarised in the diagram of Figure \IndexFigure.

Transformation matrices will belong to the 
adjoint of the Lie superalgebra
$osp(d,d|2s)$. In section \GenSuperDiffSubSection, they will be formed from
the derivative of a super-diffeomorphism parameter, \ie, from the matrix
$a_\sM{}^\sN=\*_\sM\xi^\sN$. Therefore, there is a need 
to consistently multiply matrices,
matrices and vectors, raise and lower indices and perform
transpositions. The last operation is essential in order to form the
adjoint (graded antisymmetric) from an arbitrary matrix.

Since vectors and matrices contain bosonic and fermionic objects,
ordering is important. A convention is used where covectors
and vectors transform as 
$$
\eqalign{
\delta_f W_\sM&=f_\sM{}^\sN W_\sN\komma\cr
\delta_f V^\sM&=-V^\sN f_\sN{}^\sM\cr
}\eqn
$$
under an $osp(d,d|2s)$-transformation with an algebra element $f$
(more about it later). Then $V^\sM W_\sM$ (in this order) is
invariant. Indices are always 
contracted in the direction NW-SE, and contraction is performed on
neighbouring indices.

With these conventions, indices are lowered by right multiplication
with $H$ and raised by left multiplication by a matrix $\hat H$. This
is not the inverse of $H$, but
$$
\hat H^{\sM\sN}=\left(\matrix{\eta^{\dot m\dot n}&0\cr    
  0&\epsilon^{\dot\mu\dot\nu}\cr}\right)
\eqn
$$
(taking the inverse of a matrix with this index structure is not
allowed by the conventions).

This also introduces a consistency check for the conventions for 
raising and lowering of indices, since vectors and covectors form the
same module.
Define $V^\sM=\hat H^{\sM\sN}V_\sN$. All objects dealt with have fermion
number equal to the number of ``fermionic'' indices, \ie, fundamental
$Sp(2s)$ indices. Fermion number is denoted $\e(\sM)$, which takes the
value 0 for $\sM=\dot m$ and 1 for $\sM=\dot\mu$. The invariant metric is graded
symmetric: $\hat H^{\sN\sM}=(-1)^{\e(\sM)\e(\sN)}\hat H^{\sM\sN}$. It is
convenient to define a 
super-transpose as $(M^T)^{\sM\sN}=(-1)^{\e(\sM)\e(\sN)}M^{\sN\sM}$. 

Checking the invariance of $V^\sM W_\sM=\hat H^{\sM\sN}V_\sN W_\sM$, one gets
$$
\eqalign{
\delta_f(V^\sM W_\sM)&=\hat H^{\sM\sN}(f_\sN{}^\sP V_\sP W_\sM+V_\sN
f_\sM{}^\sP W_\sP)\cr
&=(\hat H^{\sN\sP}f_\sP{}^\sM
+(-1)^{\e(\sM)\e(\sN)}\hat H^{\sM\sP}f_\sP{}^\sN)V_\sM W_\sN\cr
&=(\hat Hf+(\hat Hf)^T)^{\sN\sM}V_\sM W_\sN\komma\cr
}\eqn
$$ 
where the sign factor comes from taking $f$ and $V$ past each other,
and from transposing $\hat H$. This shows that $f$, after its first index
is raised with $\hat H$, must be graded
antisymmetric, $\hat Ha+(\hat Ha)^T=0$.
It is convenient to extend the super-transpose to matrices with the
index structure $()_\sM{}^\sN$ by
$$
(A^T)_\sM{}^\sN=(H(\hat HA)^T)_\sM{}^\sN\komma\eqn
$$
\ie,
$$
(A^T)_\sM{}^\sN=(-1)^{\e(\sM)\e(\sN)}(HA^t\hat H^t)_\sM{}^\sN
=(-1)^{(\e(\sM)+\e(\sN))\e(\sN)}(HA^t\hat H)_\sM{}^\sN\punkt\eqn
$$
Then, the superalgebra element obeys $f+f^T=0$, and
given any supermatrix $a_\sM{}^\sN$, an adjoint element is formed as
$f=a-a^T$.
Defined this way, the transpose is an
anti-involution with respect to matrix multiplication, and obeys the
standard rule 
$$
(AB)^T=B^TA^T\komma\eqn
$$
since
$$
\eqalign{
&((AB)^T)_\sM{}^\sN\cr
&=(-1)^{(\e(\sM)+\e(\sN))\e(\sN)}(H(AB)^t\hat H)_\sM{}^\sN\cr
&=(-1)^{(\e(\sM)+\e(\sN))\e(\sN)+(\e(\sM)+\e(\sP))(\e(\sP)+\e(\sN))}
          (HB^t\hat H)_\sM{}^\sP(HA^t\hat H)_\sP{}^\sN\cr
&=(-1)^{(\e(\sM)+\e(\sP))\e(\sP)+(\e(\sP)+\e(\sN))\e(\sN)}
          (HB^t\hat H)_\sM{}^\sP(HA^t\hat H)_\sP{}^\sN\cr
}\eqn
$$
(this is not a meaningful/covariant statement for other index
structures, since then the NW-SE convention is broken). There are also
statements like
$$
\eqalign{
&W_\sM=A_\sM{}^\sN V_\sN\Longleftrightarrow W^\sM=V^\sN(A^T)_\sN{}^\sM\komma\cr
&A_\sM{}^\sN=V_\sM W^\sN\Longleftrightarrow (A^T)_\sM{}^\sN=W_\sM V^\sN
\cr
}\eqn
$$
etc. The conventions lead to formulas free of extra signs due to
fermion number.

The scalar part of a matrix $M_\sM{}^\sN$ sits in its supertrace,
$$
\Str M=\tr(\hat HMH)=(-1)^{\e(\sM)}M_\sM{}^\sM=M_{\dot m}{}^{\dot m}
     -M_{\dot\mu}{}^{\dot\mu}\komma\eqn
$$
where the sign can be seen as a consequence of the NW-SE rule,
implemented in the first step.
This is because, in general, the trace of a commutator is not zero,
but acquires a sign factor due to ordering. The supertrace of a
commutator {\it is} invariant. Still, the unit matrix $\delta_\sM{}^\sN$ is of
course invariant. The singlet part of a matrix is
$$
M^{(1)}_\sM{}^\sN={1\over2d-2s}\delta_\sM{}^\sN\Str M\punkt\eqn
$$
(If $d=s$, the superalgebra $osp(2d|2d)$ is not simple,
but contains an ideal generated by the unit matrix. This will not be
relevant to our applications.)

\subsection\GenSuperDiffSubSection{Generalised super-diffeomorphisms}It is a 
straightforward exercise to define
doubled super-diffeo\-morphisms, where parameters are in the fundamental
$(2d|2s)$-dimensional module of $OSp(d,d|2s)$. The key point is the
section condition. It is already known from the modules of supersymmetry
generators under the double cover of the maximal compact subgroup that
the section condition still effectively should remove dependence on bosonic 
variables only. The supergroup provides a covariant version on
superspace, namely
$$
\hat H^{\sM\sN}\*_\sN\otimes\*_\sM=0\komma\Eqn\OSpSectionCondition
$$
where $\hat H^{\sM\sN}$ is the $OSp(d,d|2s)$-invariant metric.
Note that the section condition should be formulated on naked
derivatives, carrying indices in coordinate basis (``curved
indices''). 

The section condition (only the ``strong'' version is considered,
necessary for the algebra of generalised super-diffeomorphisms)
should be interpreted as a condition on a (maximal) linear subspace of
momenta, such that all momenta $p,p'$ in the subspace satisfy
$\hat H^{\sM\sN}p^\ms_\sN p'_\sM=0$. This amounts to finding a maximal
isotropic subspace of (co)tangent vectors. Locally, all fields depend
only on the corresponding coordinates.  
Modulo the choice of $OSp$ basis, a solution to the section condition
can always locally be brought to the form ${\*\over\*\tilde x_m}=0$, 
${\*\over\*\tilde\theta_\mu}=0$ action on all fields and parameters
(where the coordinates are defined in eq. (\InvMetricGL)).
Solutions of the $OSp$ section condition are parametrised by pure
orthosymplectic spinors, as will be described in section \RRSection.

Now, the generalised super-diffeomorphisms take the form
$$
\eqalign{
\LL_\xi V^\sM&=\xi V^\sM-V^\sN(a-a^T)_\sN{}^\sM\komma\cr
\LL_\xi V_\sM&=\xi V_\sM+(a-a^T)_\sM{}^\sN V_\sN\punkt\cr
}\eqn
$$
where $\xi=\xi^\sM\*_\sM$ and $a_\sM{}^\sN=\*_\sM\xi^\sN$ 
(the two expressions for
the transformations are of course equivalent).
A short calculation using the definitions above, and the section
condition on the form $a^Tb=0$ etc., shows that the commutator of two
super-diffeomorphisms give a new super-diffeomorphism:
$$
[\LL_\xi,\LL_\eta]=\LL_{\leftbr\xi,\eta\rightbr}\komma\eqn
$$
where
$$
\leftbr\xi,\eta\rightbr=\fr2(\LL_\xi\eta-\LL_\eta\xi)\punkt\eqn
$$
Just like for the bosonic generalised diffeomorphisms, there is a
slight violation of the Jacobi identity, taking the form of a
reducibility, related to a trivial parameter
$\zeta_\sM=\*_\sM\lambda$ with $\LL_\zeta=0$.

It has thus been shown that it is essential to have the $OSp$ structure already
at the level of ``curved'' (coordinate basis) indices, on which
generalised super-diffeomorphisms act. 
This implies that there must be a doubling not only of the bosonic
directions (as compared to physical space), but also of the fermionic
ones (as compared to ordinary superspace). Namely, if one wants to
attach engineering dimension 1 to any bosonic derivative $\*_{\dot m}$, then
it is not consistent that all fermionic derivatives have dimension
$\fr2$. Neither should one expect a formalism leading to a ``spinor
metric'' $\epsilon$ on ordinary superspace. The fermionic coordinates
here consist of $s$ coordinates $\theta^\mu$ and $s$ ``extra'' coordinates
$\tilde\theta_\mu$.
Letting $\theta$ and $\tilde\theta$ carry dimensions $-\fr2$
and $-\Fr32$ respectively is consistent with the $OSp$ structure.
Reduction to ordinary superspace is a maximal solution of the section
condition on doubled superspace. 

Having established the super-diffeomorphisms, it is straightforward to
continue with super-vielbeins, connections and curvature. The
construction parallels the one in the bosonic case, with the
orthogonal group replaced by the orthosymplectic supergroup.

\subsection\VielbeinConnectionSubSection{Compatibility, vielbein and
connection}An affine connection on superspace should be defined so that 
$$
D_\sM V_\sN=\*_\sM V_\sN+\G_{\sM\sN}{}^\sP V_\sP\eqn
$$
transforms as a tensor. This forces the connection to transform as
$$
\delta_\xi\G_{\sM\sN}{}^\sP=\LL_\xi\G_{\sM\sN}{}^\sP
  -\*_\sM(\*_\sN\xi^\sP-\xi_\sN\overleftarrow\*{}^\sP)\punkt\Eqn\GammaTransfEq
$$
The connection should, seen as matrices $\G_\sM$, take values in the Lie
superalgebra $osp(d,d|2s)$, $\G_\sM+(\G_\sM)^T=0$. This is manifestly the case
for the inhomogeneous term in eq. (\GammaTransfEq).

The entire connection comes in the tensor product of the
fundamental with the graded antisymmetric module.
Lower the last index with the invariant metric according to
$\Gamma_{\sM\sN\sP}=\Gamma_{\sM\sN}{}^\sQ H_{\sQ\sP}$.
It then follows that the totally graded antisymmetric part of the
connection transforms as a tensor, which is identified as torsion:
$$
T_{\sM\sN\sP}=-\Fr32\Gamma_{[\sM\sN\sP)}\punkt\eqn
$$

A super-vielbein $E_\sM{}^\sA$ is restricted to be a group element of
$OSp(d,d|2s)$. 
Recall that the inertial index 
$\sA=(\dot a;{\dot\alpha})=(\dot
a;\alpha,{\bar\alpha})$ 
labels representations
of the locally realised subgroup $Spin(d)\times Spin(d)$.
The inertial  index $\a$ will, in a type \II\ theory, 
label the module $s\otimes1\oplus1\otimes s$, where
$s$ is a spinor module of
$Spin(d)$. This represents the dimension-$\fr2$ directions, while the
conjugate index ${\bar\alpha}$ represents the dimension-$\Fr32$
directions\foot{A small sloppiness in index notation has been allowed 
in the use of
the same letter $\a$ for the sum of the spinor modules of the two
$Spin(d)$'s (to the right in Figure \IndexFigure) as for the spinor index of the
diagonal subgroup (the bottom of Figure \IndexFigure). There should be no danger
in this, since they label the same vector space.}.
In principle, one may imagine also a chiral situation,
with $S=s\otimes1$, and of course also extended supersymmetry with
extra R-symmetry group $R$. 
The main example will be type \II\ theory, with $d=10$ and $s=32$. 
Starting from the ``structure group'' $OSp(10,10|64)$, the locally
realised group is $Spin(1,9)\times Spin(1,9)$. 

The vielbein is demanded to satisfy a covariant constancy
(compatibility) condition,
$$
D_\sM E_\sN{}^\sA=\*_\sM E_\sN{}^\sA+\Gamma_{\sM\sN}{}^\sP E_\sP{}^\sA
-\Omega_{\sM\sB}{}^\sA E_\sN{}^\sB=0
\punkt\Eqn\SuperCompatibility
$$
Here, a spin connection on superspace, taking values in (the relevant
real form of) $so(d)\oplus
so(d)$ has been introduced. 

At this stage, it should have become obvious that, as long as only
curved indices are concerned, all considerations from double geometry
can safely be extrapolated to double supergeometry. 
For example, a 4-index curvature tensor can be formed from the affine
super-connection as
$$
\eqalign{
R_{\sM\sN\sP\sQ}&=\*_{[\sM}\Gamma_{\sN)\sP\sQ}
        +(-1)^{\e(\sN)\e(\sR)}\Gamma_{[\sM|\sP|}{}^\sR\Gamma_{\sN)\sR\sQ}\cr
&-\fr4(-1)^{(\e(\sM)+\e(\sN))\e(\sR)}\Gamma^\sR{}_{\sM\sN}\Gamma_{\sR\sP\sQ}\cr
&+(-1)^{(\e(\sM)+\e(\sN))(\e(\sP)+\e(\sQ))}(\sM\sN\leftrightarrow\sP\sQ)
\cr}\eqn
$$
(The awkward sign factors are simply a consequence of the impossibility
to write the contractions in terms of neighbouring indices).
The super-torsion Bianchi identity reads
$$
4D_{[\sM}T_{\sN\sP\sQ)}+6T_{[\sM\sN}{}^\sR
T_{|\sR|\sP\sQ)}=-3R_{[\sM\sN\sP\sQ)}
\punkt\Eqn\SuperTorsionBianchi
$$
A generalised dilaton superfield with weight 1 is introduced just as
in the bosonic case. It transforms as
$$
\delta_\xi e^{-2\Phi}=\*^\sM(\xi_\sM e^{-2\Phi})\punkt\eqn
$$
A tensor density $\Psi$ with weight $w$ transforms as
$$
\delta_\xi \Psi=\LL^{(w)}_\xi\Psi=\LL_\xi\Psi+w\*^\sM\xi_\sM\Psi\komma\eqn
$$
and its covariant derivative is
$$
D^{(w)}_\sM\Psi=D\Psi-w\Gamma^\sN{}_{\sN\sM}\Psi\punkt\eqn
$$

There are however
some differences when it comes to the interplay between the $OSp$ group
and its locally realised subgroup. 
The local subgroup is the same as in the bosonic case (but with a
different action, of course, the fermions becoming spinors), while the
torsion is considerably larger. It is straightforward to check which
modules in the spin connection (and, thereby, also in the affine
connection) remain undetermined by the compatibility condition
(\SuperCompatibility). This question is equivalent to asking which
modules appear in the spin connection, but not in the torsion. It
turns out that the undetermined part of the connection remains the same one
as in the bosonic case, the irreducible
hooks, \lower1.5pt\hbox{\yng(2,1)}, under the two $so(d)$'s, which
reside entirely in $\Omega_{\aa,\bb\cc}$. The connection components of
dimension $\fr2$ and $\Fr32$, $\Omega_{\alpha}$ and $\Omega_{\bar\alpha}$,
can be completely determined when torsion is specified. We will come
back to this correspondence when dealing with conventional constraints
in section \ConventionalConstraintsSubSection.

In the following, the procedure
of ``ordinary'' supergeometry will be mimicked. 
This means that one needs to understand
to what extent conventional transformations and constraints can be
used to eliminate parts 
of the torsion. Then, after choosing the dimension
zero torsion to a constant invariant tensor (gamma matrices), we would
like to investigate the consequences of the torsion Bianchi
identities. 

\subsection\ConventionalConstraintsSubSection{Conventional
constraints and Bianchi identities}Superfields typically have too many 
components, and some
have to be defined away by constraints. When one is dealing with a
gauge theory or geometry on superspace, there are in addition a
collection of different superfields in different representations (\eg\
components of a super-vielbein). Typically, only the lowest
dimensional field survives as independent, and effectively contain all
the higher-dimensional ones. In a geometric situation, which is based
on a super-vielbein and a spin connection, one also needs to get rid
of independent degrees of freedom in the spin connection.

The systematic way to deal with these constraints, and in particular
to make sure that they are consistent, is the method of conventional
constraints. The principles described in detail in ref. [\CGNT]. 
The consistency is ensured by the derivation of the constraints from
the use of certain transformations either redefining the vielbein by some
matrix as $E_\sM{}^\sA\rightarrow E_\sM{}^\sB M_{\sB}{}^\sA$ or
shifting the connection by some amount 
$\delta\Omega_{\sM\sA}{}^\sB=E_\sM{}^\sC\Delta_{\sC\sA}{}^\sB$. In
order to implement the constraints covariantly, one examines how the
transformations affect the torsion (so that the compatibility equation
remains to hold), and which torsion components can be set to zero or a
fixed value by this procedure. 

If the independent component superfields of the vielbein, the
connection and the torsion are listed by dimension (the vielbein is
given through a left-invariant variation
${\scr E}=E^{-1}\delta E\in osp(2d|2s)v$ in order to have flat indices and to
encode the group-valued property), one gets the superfields in Table
\FieldsTable. 
When indices are lowered on ${\scr E}$, they are graded antisymmetric.

The torsion Bianchi identities
(\SuperTorsionBianchi) will now be examined, 
starting from the lowest dimension and working
up. The treatment will not be complete, in that we will not examine
all the irreducible modules of torsion and Bianchi identities.

First, note that there is no torsion below dimension $-\fr2$, and
thus no conventional constraint that can remove the vielbein at
dimension $-1$. A linearised field ${\scr E}_{(\alpha\beta)}$ will indeed
be the superfield containing all fields (see section \FieldsSection).

\Table\FieldsTable
{\vtop{
$$
\EqMatrix{\hfill\hbox{dim}&\hfill-1&{\scr E}_\alpha{}^{\bar\beta}&&\cr
&\hfill-\fr2&{\scr E}_\alpha{}^\bb\sim{\scr E}_\aa{}^{\bar\beta}
                   &&T_{\alpha\beta\gamma}\cr
&\hfill0&\qquad{\scr E}_\aa{}^\bb,\,{\scr E}_\alpha{}^\beta\sim
{\scr E}_{\bar\alpha}{}^{\bar\beta}\qquad
      &&T_{\alpha\beta\cc}\cr
&\hfill\fr2&{\scr E}_{\bar\alpha}{}^\bb\sim{\scr E}_\aa{}^\beta
                  &\Omega_\alpha
                  &\qquad T_{\alpha\beta\bar\gamma}\,,\,T_{\alpha\bb\cc}\qquad\cr
&\hfill1&{\scr E}_{\bar\alpha}{}^\beta&\Omega_\aa
               &T_{\aa\bb\cc}\,,\,T_{\alpha\bb\bar\gamma}\cr
&\hfill\Fr32&&\Omega_{\bar\alpha}&T_{\alpha\bar\beta\bar\gamma}\,,\,T_{\aa\bb\bar\gamma}\cr
&\hfill2&&&T_{\aa\bar\beta\bar\gamma}\cr
&\hfill\Fr52&&&T_{\bar\alpha\bar\beta\bar\gamma}\cr
}
$$
}}
{Vielbein, spin connection and torsion superfields.}

At dimension $-\fr2$ there is completely symmetric torsion
$T_{\alpha\beta\gamma}$. Remember that the index $\alpha$ downstairs denotes
collectively describes two spinors,
$$
\alpha \leftrightarrow ({\bf16},{\bf1})\oplus({\bf1},{\bf16})
=(00000)(00010)\oplus(00010)(00000)\komma\eqn
$$ 
of $Spin(1,9)\times Spin(1,9)$. Conventional constraints may be
imposed corresponding to ${\scr E}_{\alpha\dot b}$.
It now turns out that not all torsion can be removed by conventional
constraints. The remaining torsion is in the module
$$
T_{-{1\over2}}=(00000)(00030)\oplus(00010)(00020)
        \oplus(00020)(00010)\oplus(00030)(00000)\punkt
\Eqn\DimMinusHalfPhysical
$$
This should be recognised as the modules occurring in the expansion of
a scalar field depending on a pair of pure spinors
$\lambda^\alpha=(\lambda^{\alpha'},\lambda^{\alpha''})$, one for each
$Spin(1,9)$, to third order. Setting the corresponding torsion
components to 0 will be a physical, not conventional, constraint, and
will give rise to the cohomology described in section \FieldsSection. For
now we will proceed by setting the entire torsion at dimension $-\fr2$
to $0$. This physical constraint will propagate through the superfield
and force the theory on shell.

At dimension $0$, the torsion is $T_{\alpha\beta\dot c}$ and there are
conventional constraints corresponding to ${\scr E}_{\dot a\dot b}$
and ${\scr E}_{\alpha\bar\beta}$. A large number of components remain
in the torsion after the conventional constraints are exhausted. 
There is however a Bianchi identity at dimension $0$, with the index
structure $(\alpha\beta\gamma\delta)$, where no curvature
participates. This gives a number of additional constraints on the
torsion. A na\"\i ve counting ``torsion minus vielbein minus Bianchi''
indicates that no torsion survives. 
It is consistent with the Bianchi identity, thanks to the usual
10-dimensional Fierz identity 
$\gamma_{a'(\alpha'\beta'}\gamma^{a'}{}_{\gamma'\delta')}=0$, to set the torsion to
$$
\eqalign{
T_{\alpha'\beta'}{}^{c'}&=2\gamma_{\alpha'\beta'}^{c'}\komma\cr
T_{\alpha''\beta''}{}^{c''}&=2\gamma_{\alpha''\beta''}^{c''}\komma\cr
}\eqn
$$
for which the shorthand
$T_{\alpha\beta}{}^{\dot c}=2\gamma_{\alpha\beta}{}^{\dot c}$ 
is introduced. A
non-zero torsion at dimension $0$ is needed both to establish a relation to
conventional superspace and to remove torsion through the vielbein
part of conventional
constraints at all dimensions.

Note that what has occurred in the analysis this far is quite
different from ordinary superspace. There, the lowest-dimensional part
of the vielbein, containing all physical fields, comes at dimension
$-\fr2$, and physical constraints are implemented in the torsion at
dimension $0$. This difference will be given a natural interpretation
in section \FieldsSection.

At dimension $\fr2$, the fields that could possibly appear in the
torsion are spinors. In supersymmetric double field theory, this is
however not expected. There are vector-spinor gravitini in
$({\bf10},\overline{\bf16})\oplus(\overline{\bf16},{\bf10})$ and spinors in
$({\bf1},{\bf16})\oplus({\bf16},{\bf1})$. However, the ``generalised
gravitino'' consists of both together, in the sense that also the
spinors transform inhomogeneously (with a derivative on the parameter) 
under local supersymmetry [\HohmKwak,\JeonLeeParkIII]. So none
of these fields are allowed to appear in torsion, which is covariant. 
There are 3 spinors (under each $Spin(1,9)$) in
$T_{\alpha\beta\bar\gamma}$ and one in $T_{\alpha\dot b\dot c}$. The
latter can be set to zero using $\Omega_\alpha$. The Bianchi identity
(there is no curvature at this dimension)
contains two spinors, and it can be checked that nothing survives that
can not be absorbed by a conventional constraint, which agrees with
the expectations.
Concerning the rest of the modules appearing the the dimension $\fr2$ 
torsion, a detailed analysis has not been performed, but counting
indicates that
conventional constraints and Bianchi identities remove everything.
It was noted earlier that the only part of the connection that remains
undetermined, using the vielbein compatibility and fixing the torsion,
is the same as in bosonic double geometry. Having vanishing torsion at
dimension $\fr2$ completely determined $\Omega_\alpha$ in terms of the
vielbein.

Moving to dimension 1, this is where, in conventional supergeometry,
field strengths for tensor fields typically 
appear in the torsion (and curvature). The
Ramond-Ramond field strengths form a $Spin(d,d)$ spinor, which in flat
indices becomes a field $F^{\alpha'\beta''}$ (after self-duality is
imposed). Here, unlike in ordinary superspace, the dimension 1
vielbein can be used to absorb the Ramond-Ramond field strength, which becomes
geometric (this is also observed in
refs. [\HatsudaKamimuraSiegelI,\HatsudaKamimuraSiegelII]). 
Effectively, by the conventional constraint, the separate superfield
in the dimension $1$ vielbein is eliminated, since the corresponding
degrees of freedom already occur in the $\theta$ expansion of the
dimension $-1$ vielbein.

At dimension $\Fr32$, there are also more conventional constraints
available than usual, and the gravitino field strength in 
can be removed
from the torsion by invoking the conventional constraints
corresponding to the spin connection $\Omega_{\bar\alpha}$. (Its
integrability should then instead appear as a Bianchi identity 
$R_{[\aa\bb\cc]\bar\delta}=0$ at dimension $\Fr52$.)

Torsion exists {\it a priori} up to dimension $\Fr52$.
We would of course like also the higher torsion components to vanish, 
in order not to produce any fields outside the supergravity.
The Bianchi identities seem to ensure this (the
detailed identities for all modules have not been examined, 
but a counting supports the claim).

At dimension 2, one should find the equations of motion for the double
geometry, as well as the equations of motion (equivalently, Bianchi
identities) for the RR field strength. 
We have not performed the complete calculation at dimension 2, but
expect the Bianchi identities for the (vanishing) torsion to contain
all information. 

Consider for example the Ramond-Ramond equations of motion, which may
appear in the modules $(00001)(00010)\oplus(00010)(00001)$. This may
come in the torsion as
$$
T_{a'}{}^{\alpha'\alpha''}=\gamma_{a'}{}^{\alpha'\beta'}K_{\beta'}{}^{\alpha''}
\komma\eqn
$$
and the corresponding expression with the $Spin(1,9)$'s exchanged.
It is forced to zero by the Bianchi identity with indices
$a''b'\gamma''\bar\delta'$. Then, the Bianchi identity with indices 
$a'b'\gamma'\bar\delta''$ ensures that it also vanishes in the curvature.

One may also check for the equations of motion of the double geometry,
in the linearised coset representative $(10000)(10000)$. It is of
course not implied by the purely bosonic Bianchi identity,
$R_{[\aa\bb\cc\dd]}=0$, but comes in
$R_{a'b'd''}{}^{c'}=\delta_{[a'}{}^{c'}S_{b']d''}$ and in $R_{a''b''d'}{}^{c''}$.
Now there is also a Bianchi identity with structure
$\aa\bb\gamma\bar\delta$. The part of this curvature where the spinor
pair of indices, but not the vector pair, is $so(1,9)\oplus
so(1,9)$-valued is identified with the corresponding part of the 
curvature with vector indices above,
$$
R_{a'b''\gamma'}{}^{\delta'}
=\fr4(\gamma^{c'd'})_{\gamma'}{}^{\delta'}R_{a'b''c'd'}
=-\fr4(\gamma_{a'}{}^{c'}){}_{\gamma'}{}^{\delta'}S_{c'b''}\Eqn\EOMCurvature
$$
etc.
The torsion may also contain a term 
$T_{a''}{}^{\beta'\gamma'}\sim\gamma^{e'\beta'\gamma'}S_{e'a''}$. It
contributes to the Bianchi identity with indices 
$a'b''\gamma'\bar\delta'$ with a term proportional to the curvature
in eq. (\EOMCurvature), but also with a term containing
$\delta_{\gamma'}{}^{\delta'}S_{a'b''}$. This term must thus vanish both in
torsion and curvature.

The results of the geometric analysis here are supported by the
examination of the cohomology of the lowest-dimensional vielbein
component performed in section \FieldsSection.

\section\FieldsSection{Fields from pure spinor cohomology}The full
implementation of the conventional constraints implies that all
geometric superfields can be expressed in terms of the
lowest-dimensional part of the vielbein with dimension $-1$. In section 
\ConventionalConstraintsSubSection\ it was noted that already the
lowest-dimension Bianchi identity, at dimension $-\fr2$, involving the
torsion $T_{\alpha\beta\gamma}$, could not be completely eliminated,
unless some physical constraints (\DimMinusHalfPhysical) were
imposed, which happen to coincide with the expansion to third order of
a scalar field depending on a pair of pure $Spin(1,9)$ spinors
$\Lambda=(\lambda',\lambda'')$. 

This seems to indicate that the double supergeometry may be encoded in the
framework of pure spinor superfields
[\BerkovitsI\skipref\CederwallNilssonTsimpisI\skipref\CederwallNilssonTsimpisII\skipref\BerkovitsParticle\skipref\SpinorialCohomology\skipref\CederwallBLG\skipref\CederwallABJM\skipref\PureSGI-\PureSGII],
see ref. [\PureSpinorOverview] for an overview. In
order to examine this hypothesis, it is convenient to work at a
linearised level. The linearised dimension $-1$ vielbein 
${\scr E}_{\alpha\beta}$ should appear at second order in the
expansion in $\Lambda$. A BRST operator is
$$
{\scr
Q}=Q'+Q''=\lambda^{\alpha'}D_{\alpha'}+\lambda^{\alpha''}D_{\alpha''}
\punkt\eqn
$$
The field content is obtained from the zero-mode cohomology, which
will be the tensor product of the zero-mode cohomologies of $Q'$ and
$Q''$. It is well known that each of them contain a $D=10$
super-Yang--Mills cohomology, including ghosts and antifields. The SYM
cohomology is listed in Table \TableOne.
The double cohomology is then obtained as the tensor products of two
SYM cohomologies (labelled by modules in the two $Spin(1,9)$
groups). It is listed in Table \TableTwo. Both tables are organised so that
each column contains the $\theta$ expansion of the superfield
occurring at a certain power of the pure spinors. The superfields have
been shifted down in order to display components of the same dimension
on the same row. A ``$\bullet$'' indicates the absence of zero-mode cohomology.

Let us take a closer look at the contents of Table \TableTwo. Its lowest
component is a scalar with ghost number $2$. It represents the singlet
reducibility of the super-diffeomorphisms (which, in turn, is inherited
from the $B$ field). At ghost number 1, one finds the vector and the two
spinors of the super-diffeomorphisms, decomposed in $Spin(1,9)\times
Spin(1,9)$ modules. At ghost number $0$, the physical fields
appear. At dimension $0$, there is the linearised coset module of the
double geometry. At dimension $\fr2$, one finds the correct modules 
(spinor and vector-spinor) for the generalised gravitino potential. 
At dimension 1, there is a vector, the dilaton field strength,
together with a bispinor, the Ramond-Ramond field strength. It should
be noted that the RR fields (as usual) only enter the supergeometry
through their field strength. There is no way of accommodating the
potentials together with their gauge symmetry in a geometric
framework, where the gauge symmetry consists of
super-diffeomorphisms. A full treatment of RR fields and gauge
transformations will be given in the following section.
Continuing to ghost number $-1$, the linearised
equations of motion are read off, 
and one finds the correct modules at dimension $\Fr32$
and $2$. There is however a doubling, due to the section
condition. Roughly speaking, $(\*')^2$ and $(\*'')^2$ contribute
independently, and their sum and difference give the section condition
and the equations of motion. The details of this are left for possible
future examination.

We conclude that the double pure spinor cohomology represents the
physical fields of the double supergeometry. This interpretation is
so far only at linearised level around \eg\ a flat superspace, and it
is not geometric, since all vielbein components except the one at
dimension $-1$ have been discarded. Note that the zero-mode
cohomology of Table \TableTwo\ (unlike the one in Table \TableOne, and \eg\ the
cohomology of $D=11$ supergravity [\PureSGI,\PureSGII]) 
does not exhibit a field-antifield symmetry. This is normal in a
situation including self-dual fields, and prevents the existence of a
pure spinor 
superfield action. In principle, interactions may be introduces as
deformations of the cohomology, but there is no full
Batalin--Vilkovisky pure spinor superfield framework.

\Table\TableOne{
$$
\vtop{\baselineskip20pt\lineskip0pt
\ialign{
$\hfill#\quad$&$\,\hfill#\hfill\,$&$\,\hfill#\hfill\,$&$\,\hfill#\hfill\,$
&$\,\hfill#\hfill\,$&$\,\hfill#\hfill$&\qquad\qquad#\cr
\hbox{ghost} \#= &1 &0    &-1     &-2    &-3\cr
\hbox{dim}=0&(00000)&       &       &       &\phantom{(00000)}       \cr
        \fr2&\bullet&\bullet&               &       \cr 
           1&\bullet&(10000)&\bullet&       &       \cr
       \Fr32&\bullet&(00001)&\bullet&\bullet&       \cr
           2&\bullet&\bullet&\bullet&\bullet&\bullet\cr
       \Fr52&\bullet&\bullet&(00010)&\bullet&\bullet\cr
           3&\bullet&\bullet&(10000)&\bullet&\bullet\cr
       \Fr72&\bullet&\bullet&\bullet&\bullet&\bullet\cr
           4&\bullet&\bullet&\bullet&(00000)&\bullet\cr
       \Fr92&\bullet&\bullet&\bullet&\bullet&\bullet\cr
}}
$$
}
{The zero-mode cohomology of 
$D=10$ super-Yang--Mills.}

\Table\TableTwo{\sixpoint
$$
\vtop{\baselineskip24pt\lineskip0pt
\ialign{
$\hfill#\quad$&$\hfill#\hfill$&$\hfill#\hfill$&$\hfill#\hfill$
&$\hfill#\hfill$&$\hfill#\hfill$&$\hfill#\hfill$&$\hfill#\hfill$
        \cr
\hbox{\xrm gh} \#= &2 &1    &0     &-1    &-2&-3&-4\cr
\hbox{\xrm dim}=-2&(00000)(00000)&    &  & &      \cr
        -\Fr32&\bullet&\bullet&               &       \cr 
           -1&\bullet&\raise4pt\vtop{\baselineskip6pt\ialign{
					\hfill$#$\hfill\cr
					(00000)(10000)\cr
					(10000)(00000)\cr}}
                        &\bullet&       &       \cr
       -\fr2&\bullet&\raise4pt\vtop{\baselineskip6pt\ialign{
					\hfill$#$\hfill\cr
					(00000)(00001)\cr
					(00001)(00000)\cr}}&\bullet&\bullet& 
                                                                   \cr
           0&\bullet&\bullet&(10000)(10000)&\bullet&\bullet\cr
       \fr2&\bullet&\bullet&\raise8pt\vtop{\baselineskip6pt\ialign{
					\hfill$#$\hfill\cr
					(00000)(00010)\cr
                                        (00001)(10000)\cr
                                        (00010)(00000)\cr
                                        (10000)(00001)\cr}}
                                        &\bullet&\bullet&\bullet\cr
           1&\bullet&\bullet&\raise6pt\vtop{\baselineskip6pt\ialign{
					\hfill$#$\hfill\cr
					(00000)(10000)\cr
                                        (00001)(00001)\cr
                                        (10000)(00000)\cr}}
                                        &\bullet&\bullet&\bullet&\bullet\cr
       \Fr32&\bullet&\bullet&\bullet&
                \raise4pt\vtop{\baselineskip5pt\ialign{
					\hfill$#$\hfill\cr
					(00010)(10000)\cr
                                        (10000)(00010)\cr}}
                                       &\bullet&\bullet&\bullet\cr
           2&\bullet&\bullet&\bullet&
                        \raise8pt\vtop{\baselineskip6pt\ialign{
			\hfill$#$\hfill\cr
                   		2(00000)(00000)\cr
                                        (00001)(00010)\cr
                                        (00010)(00001)\cr
                            2(10000)(10000)\cr}}&
                            \bullet&\bullet&\bullet\cr
       \Fr52&\bullet&\bullet&\bullet&
                \raise4pt\vtop{\baselineskip5pt\ialign{
					\hfill$#$\hfill\cr
					(00001)(10000)\cr
                                        (10000)(00001)\cr}}
                                &\bullet&\bullet&\bullet\cr
       3&\bullet&\bullet&\bullet&\bullet
                        &\raise6pt\vtop{\baselineskip6pt\ialign{
					\hfill$#$\hfill\cr
					(00000)(10000)\cr
                                        (00010)(00010)\cr
                                        (10000)(00000)\cr}}
                                       &\bullet&\bullet\cr
       \Fr{7}2&\bullet&\bullet&\bullet&\bullet
                        &\raise8pt\vtop{\baselineskip6pt\ialign{
					\hfill$#$\hfill\cr
					(00000)(00001)\cr
                                        (00010)(10000)\cr
                                        (00001)(00000)\cr
                                        (10000)(00010)\cr}}
                                        &\bullet&\bullet\cr
       4&\bullet&\bullet&\bullet&\bullet&(10000)(10000)
                                &\bullet&\bullet\cr
       \Fr{9}2&\bullet&\bullet&\bullet&\bullet&\bullet
                        &\raise4pt\vtop{\baselineskip6pt\ialign{
					\hfill$#$\hfill\cr
					(00000)(00010)\cr
					(00010)(00000)\cr}}
                                        &\bullet\cr
       5&\bullet&\bullet&\bullet&\bullet&\bullet
                        &\raise4pt\vtop{\baselineskip6pt\ialign{
					\hfill$#$\hfill\cr
					(00000)(10000)\cr
					(10000)(00000)\cr}}
                                        &\bullet\cr
       \Fr{11}2&\bullet&\bullet&\bullet&\bullet&\bullet&\bullet&\bullet\cr
       6&\bullet&\bullet&\bullet&\bullet&\bullet&\bullet&(00000)(00000)
                \cr
}}
$$
}
{Zero-mode cohomology of 
supersymmetric double field theory as $(\hbox{SYM})^2$}

\vfill\eject

\section\RRSection{$OSp$ spinors and Ramond-Ramond fields}In this
section, 
a basis for the the $OSp(d,d|2s)$ vector
representation is used such that
$dZ^\sM=(dx^m,d\tilde x_m;d\theta^\mu,d\tilde\theta_\mu)$. Here,
$M=(m,\mu)$ is a $GL(d|s)$ index, and the invariant metric takes the
form (\InvMetricGL).

\subsection\OSpSpinorsSubSection{$OSp$ spinors}In complete 
analogy with $Spin(d,d)$, where a spinor is realised as a
sum of even or odd forms in $d$ dimensions,
we now form a basis for the infinite-dimensional spinor representations of
$osp(d,d|2s)$ as consisting of all superforms 
$$
e^{m_1\ldots m_p|\mu_1\ldots\mu_q}
=dx^{m_1}\wedge\ldots\wedge dx^{m_p}\wedge 
d\theta^{\mu_1}\wedge\ldots\wedge d\theta^{\mu_q}\punkt\eqn
$$
Chiral spinors have $p+q$ even or odd. 
Ramond-Ramond superfields are even or odd forms on ordinary
superspace. Fields in the two representations are
$$
\eqalign{
S&=\bigoplus\limits_{P\in2\NN}\fr{p!}dz^{M_1}\wedge\ldots\wedge
dz^{M_P}S_{M_P\ldots M_1}\komma\cr
C&=\bigoplus\limits_{P\in2\NN+1}\fr{p!}dz^{M_1}\wedge\ldots\wedge
dz^{M_P}C_{M_P\ldots M_1}\punkt\cr        
}\eqn
$$
Here, the conventions for ordering and contractions of
section \OSpBasicsSubSection\ lead to the standard conventions for
forms on superspace.

``Super-Gamma matrices'' $\Sigma^\sM$ are now introduced through their action 
by wedge products and contractions on the
``spinor'' superforms: 
$$
\matrix{\Sigma^M\omega=\sqrt2dz^M\wedge\omega\,\,:\hfill
&\left\{\matrix{\Sigma^m\omega=\sqrt2dx^m\wedge\omega\cr
               \Sigma^\mu\omega=\sqrt2d\theta^\mu\wedge\omega}\right.
                                        \hfill\cr
&\cr
\Sigma_M\omega=\sqrt2\imath_M\omega\,\,:\hfill
&\left\{\matrix{ \Sigma_m\omega=\sqrt2\imath_m\omega\cr
                \Sigma_\mu\omega=\sqrt2\imath_\mu\omega}\right.
                                        \hfill\cr
}\Eqn\SigmaMatrixDef
$$
This definition immediately leads to
$$
\{\Sigma^\sM,\Sigma^\sN]=2\hat H^{\sM\sN}\id\komma\eqn
$$
where $\{A,B]=AB+(-1)^{|A||B|}BA$ is the graded anticommutator.
Orthosymplectic transformations are realised as
$$
\delta_f\omega=-\fr4\Sigma^{\sM\sN}f_{\sN\sM}\omega\komma\eqn
$$
with $\Sigma^{\sM\sN}=\Sigma^{[\sM}\Sigma^{\sN)}$ graded antisymmetric.

The two spinor chiralities form infinite-dimensional highest weight 
modules $S$ and $C$ of (the
double cover of) $OSp(d,d|2s)$. Their conjugates modules are lowest
weight, and the only singlets appearing in tensor products of spinors
and cospinors are in $S\otimes\bar S$ and $C\otimes\bar C$ (unlike the
$Spin(d,d)$ situation where the conjugate module $\bar S$ is $S$ or
$C$, depending on dimension).
Note that action with a single $\Sigma$ matrix changes the chirality
of the spinor.

\subsection\RRSpinorSubSection{Ramond-Ramond superfields}The treatment of
Ramond-Ramond fields as dynamical spinors in double field theory was
introduced in ref. [\HohmKwakZwiebachI]. Here, it is extended to
superspace. In the supergeometry, only the RR field strengths
appeared, since their gauge symmetry could not be encoded. This will
now be remedied.

Introduce a Dirac 
operator $\dslash=\Sigma^\sM\*_\sM$, mapping $S$ to $C$
and vice versa. It becomes nilpotent thanks to the section condition
(\OSpSectionCondition):
$$
\dslash^2\omega=\Sigma^\sM\*_\sM\Sigma^\sN\*_\sN\omega
=\Sigma^\sM\Sigma^\sN\*_\sN\*_\sM\omega=2\hat H^{\sM\sN}\*_\sN\*_\sM\omega=0
\punkt\eqn
$$
It is therefore meaningful to let a gauge field (RR superpotential)
transform in $S$, say. Let us call this field $S$. It will have a field
strength $F=\dslash S$ which is invariant under the gauge
transformations\foot{These gauge transformations are infinitely
reducible, with spinors all the way down. 
This happens already for the $Spin(d,d)$ RR-fields in
double geometry. There, the naive counting $1-1+1-1+\ldots=\fr2$ gives the
correct counting of the off-shell degrees of freedom modulo gauge
transformations.}  $\,\delta S=\dslash\Lambda$. 

We must however check if the Dirac operator is covariant. This is done
by replacing the naked derivative with a covariant one,
$\dslash\rightarrow\Dslash$, containing the affine super-connection
$\Gamma$. One can also allow for a weight $w$.
$$
\eqalign{
\Dslash\omega&=\Sigma^\sM D_\sM\omega=\Sigma^\sM\left(\*_\sM
-\fr4\Gamma_\sM{}^{\sN\sP}\Sigma_{\sP\sN}
-w\Gamma^\sN{}_{\sN\sM}\right)\omega\cr
&=\left(\dslash-\fr4\Gamma^{\sM\sN\sP}\Sigma_{\sP\sN\sM}
        -(w-\fr2)\Gamma^\sN{}_\sN{}^\sM\Sigma_\sM\right)\omega\punkt\cr
}\eqn
$$
As for the $Spin(d,d)$ spinor, the $\Sigma^{(3)}$ term is torsion, and
the (naked) Dirac operator becomes covariant if\foot{We have not been
able to trace this statement or its derivation 
in the literature. It seems to be taken
for granted in ref. [\HohmKwakZwiebachI].} $\,w=\fr2$, in which
case it becomes
$$
\dslash\omega=(\Dslash-\fr6T^{\sA\sB\sC}\Sigma_{\sC\sB\sA})\omega
\punkt\Eqn\CovDerOnSpinor
$$
In a suitable basis adjusted to the solution of the section
condition, it is the exterior derivative on superspace forms. Take
only $T_{\a\b}{}^c=2\g_{\a\b}{}^c$ non-vanishing. In a flat basis the
$OSp$ spinor is a form spanned by $E^a$ and $E^\alpha$. The torsion
term in eq. (\CovDerOnSpinor) becomes
$-E^\beta\wedge E^\alpha\gamma_{\alpha\beta}{}^c\imath_c$,
corresponding to the standard dimension 0 torsion on ordinary
superspace. 
This leads to the usual dimension $0$ Chevalley--Eilenberg
cohomologies [\AzcarragaTownsend] for the Ramond-Ramond superfields,
listed for example in 
ref. [\DBranesII].

It should be expected that the Ramond-Ramond double superfield encodes
also the supergeometric data, \ie, all the fields in double
supergeometry, discussed in section \DoubleSuperGeometrySection. This
would be in accordance with \eg\ $D=11$ supergravity, where the true
basic superfield is the one corresponding to the 3-form tensor
[\PureSGI,\PureSGII]. We
have not checked how this happens, but leave it for future work. Due
to the infinite reducibility of the gauge transformation, it is not
clear what kind of structure will replace the pure spinor superfields
of section \FieldsSection. Whether such a structure can be used for a
Batalin--Vilkovisky action (or pseudo-action, remembering the
selfduality), or some similar efficient description of the full
dynamics, remains to be seen.

\subsection\PureSpinorSubSection{Pure OSp spinors and
super-sections}Just as pure spinors define isotropic subspaces ---
sections --- in double field theory, a pure $OSp$ spinor 
defines a super-section. This is an isotropic embedding of an ordinary
$(d|s)$-dimensional superspace in the $(2d|2s)$-dimensional doubled
superspace. Since the spinors are infinite-dimensional, there is no
analogue of a Mukai pairing, and it is more convenient to use the
traditional definition of a pure spinor due to Cartan [\CartanSpinors]
than to form spinor bilinears. 

A pure spinor is a spinor that is
annihilated by a maximal isotropic set of
$\Sigma$-matrices. Inspecting eq. (\SigmaMatrixDef), it is clear that
the spinor represented by $1$ has this property; it is annihilated by 
$\Sigma_M=\imath_M$. Such a pure spinor lies in a minimal
(finite-dimensional) orbit of the
(infinite-dimensional) $S$ module under $OSp(d,d|2s)$. Acting with the
supergroup generates the orbit
$$
\Lambda=e^{\Phi+B\wedge}1\komma\Eqn\PureSpinorOrbit
$$
where $B$ is a super-2-form in $(d|s)$ dimensions.
The pure spinor space is the supergroup quotient
$$
\Pi={OSp(d,d|2s)\over GL(d|s)\ltimes{{\scr B}}}\times\RR\komma\eqn
$$
where ${\scr B}$ is the graded antisymmetric module. The
dimensionality of pure spinor space is 
$$
\hbox{dim}(\Pi)=\bigl(\,\Fr{d(d-1)}2+\Fr{s(s+1)}2+1\,\big|\,ds\,\bigr)
\komma\eqn
$$
which clearly matches the dimensionality of the orbit in
eq. (\PureSpinorOrbit). 

The pure $OSp$ spinors should be relevant for the formulation of
D-brane dynamics in double superspace, much in the same way pure
spinors enter the construction of D-branes in (bosonic) double field
theory [\AsakawaSasaWatamura,\MaDBrane,\BermanCederwallMalek]. 
The D-brane, like
the section, is a maximal isotropic subspace [\HullT].

\section\ConclusionSection{Conclusions}We have constructed a double
supergeometry, where covariance under super-diffeomorphisms is
manifest. Ordinary superspace is obtained as a super-section, just as
ordinary space is a section in double geometry. In a maximally
supersymmetric situation, the fields will be on shell, when a set of
physical constraints (in contrast to the conventional ones) are
imposed at the lowest-dimensional torsion. The structure is reflected
in the product of two super-Yang--Mills pure spinor cohomologies.

A few comments on the relation to the work by Hatsuda \etal\ 
[\HatsudaKamimuraSiegelI,\HatsudaKamimuraSiegelII] are in place. That
work starts from affine super-Poincar\'e algebras for left- and
right-movers on a string, which leads to an orthosymplectic group of
significantly higher dimension than the one in the present paper. Then
$\kappa$-symmetry and Virasoro symmetry are imposed in order to
restrict the background. There is only torsion, no curvature, but
what in the present work is curvature is encoded as part of the
torsion (since what we here call spin connection is made
part of a big vielbein). A
treatment of super-diffeomorphisms is not performed. In many respects,
the results concerning supergravity fields of the present work and of
refs. [\HatsudaKamimuraSiegelI,\HatsudaKamimuraSiegelII], such as the
appearance of a ``prepotential'', the vielbein at dimension $-1$,
seem to be consistent, and it is likely that they are equivalent.

A natural question is how this can be continued to exceptional field
theory. Unlike the orthogonal group $O(d,d)$, the exceptional duality
groups $E_{n(n)}$, have no finite-dimensional super-extensions
[\KacSuperalgebras]. The present work may offer some clues. On one
hand it is well known that the coordinate module in exceptional
geometry contains a spinor when the U-duality group $E_{n(n)}$ is
reduced to the T-duality group $Spin(n-1,n-1)$. On the other hand we
have seen in section \RRSection\ that superspace counterpart of  
the T-duality spinor is an infinite-dimensional module. It seems clear
that it will be necessary to start from a superspace with an
infinite-dimensional coordinate module, and since there are no
finite-dimensional superalgebras at hand, also the super-extension of
the U-duality group should be infinite-dimensional, maybe of the type
depicted in Figure \DynkinFigure. 
Examination of this hypothesis will be the subject of future work.

\Figure\DynkinFigure{\epsffile{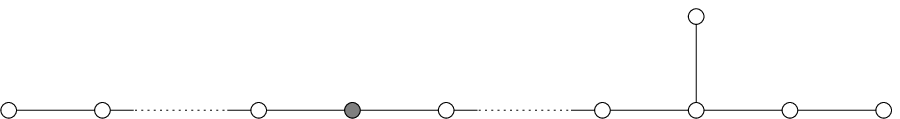}}
{The Dynkin diagram for a
superalgebra in exceptional supergeometry?}

\acknowledgements{The author would like to extend his gratitude to 
Machiko Hatsuda for discussions and explanation of her work, and to
Jakob Palmkvist for clarifying some aspects of superalgebras.}

\refout

\end